\documentclass[a4paper,11pt]{article}

\pdfoutput=1 % if your are submitting a pdflatex (i.e. if you have
\usepackage{jheppub} % for details on the use of the package, please
\usepackage[dvipsnames]{xcolor}
\usepackage{tikz}
\usetikzlibrary{external}
\usetikzlibrary{patterns,arrows,decorations.pathmorphing,decorations.markings}
\usepackage{slashed,adjustbox,bm,mathrsfs,bbm}
\usepackage{subcaption}
\usepackage{cleveref}
\usepackage[T1]{fontenc} % if needed
\usepackage[utf8]{inputenc}
\usepackage{mathtools}
\usepackage{float}
\usepackage{amsmath}
\usepackage{soul}
\allowdisplaybreaks

\newcommand{\nn}{\nonumber \\}

\newcommand{\cM}{\mathcal{M}}
\newcommand{\cT}{\mathcal{T}}

\newcommand{\cD}{\mathcal{D}}
\newcommand{\cL}{\mathcal{L}}

\newcommand{\cY}{\mathcal{Y}}

\def\rd{{\rm d}}

\def\nn{ \nonumber \\ }
\def\coef#1#2#3{ \;{\vphantom{A^2}}^{#1}\!{{C}}^{#3}_{#2} }
\def\dcoef#1#2#3{ {\vphantom{A^2}}^{#1}\!{\dot{{C}}}^{#3}_{#2} }

\def\op#1#2#3{ {\vphantom{A^2}}^{#1}\!{Q}^{#3}_{#2} }

\newcommand{\kappaHypKinEight}{\kappa_{1}^{(8)}}
\newcommand{\kappaPauliKinEight}{\kappa_{2}^{(8)}}
\newcommand{\kappaPauliKinEightTwo}{\kappa_{3}}
\newcommand{\kappaPauliKinEightFour}{\kappa_{\epsilon}}

\newcommand{\kappaYukawaFourYukawaEightPrime}{\kappa_{7}^{\prime(8)}}

\newcommand{\kappaYukawaKinEight}{\kappa_{9}^{(8)}}

\newcommand{\kappaYukawaKinThreeEight}{\kappa_{10}^{(8)}}

\newcommand{\kappaYukawaKinTwoEight}{\kappa_{12}}

\newcommand{\dimEightBosonLoopOne}{\theta_{1}}
\newcommand{\dimEightBosonLoopTwo}{\theta_{2}}
\newcommand{\dimEightBosonLoopThree}{\theta_{3}}
\newcommand{\dimEightBosonLoopYuk}{\theta_{4}}
\makeatletter
\newsavebox{\@brx}
\newcommand{\llangle}[1][]{\savebox{\@brx}{\(\m@th{#1\langle}\)}%
  \mathopen{\copy\@brx\kern-0.5\wd\@brx\usebox{\@brx}}}
\newcommand{\rrangle}[1][]{\savebox{\@brx}{\(\m@th{#1\rangle}\)}%
  \mathclose{\copy\@brx\kern-0.5\wd\@brx\usebox{\@brx}}}
\makeatother

\subheader{\normalsize
	\vspace*{-3.7em}
	\begin{flushright}
    CERN-TH-2025-084\\
	\end{flushright}
}

\title{\boldmath Renormalizing Two-Fermion Operators in the SMEFT via Supergeometry}

\author[a,b]{Benoît Assi,}

\author[c]{Andreas Helset,}

\author[d]{Julie Pag\`es,}

\author[e,f,g]{and Chia-Hsien Shen}

\affiliation[a]{Theory Division, Fermilab, Batavia, IL 60510, USA}
\affiliation[b]{Department of Physics, University of Cincinnati, Cincinnati, OH 45221, USA}

\affiliation[c]{Theoretical Physics Department, CERN, 1211 Geneva 23, Switzerland}

\affiliation[d]{Physics Department,
University of California San Diego, 9500 Gilman Drive,\\ La Jolla, CA 92093-0319, USA}

\affiliation[e]{
Department of Physics and Center for Theoretical Physics,
National Taiwan University, Taipei 10617, Taiwan
}
\affiliation[f]{
Leung Center for Cosmology and Particle Astrophysics, Taipei 10617, Taiwan}
\affiliation[g]{
Physics Division, National Center for Theoretical Sciences, Taipei 10617, Taiwan}

\emailAdd{bassi@fnal.gov}
\emailAdd{andreas.helset@cern.ch}
\emailAdd{jcpages@ucsd.edu}
\emailAdd{chshen@phys.ntu.edu.tw}

\abstract{
We extend the geometric framework of field-space covariance for loop computations, thereby unifying the treatment of scalars, fermions, and gauge bosons in effective field theories. This allows us to derive a manifestly covariant formula for one-loop UV divergences that includes contributions from mixed boson-fermion graphs. The result is expressed in terms of geometric invariants of the field-space supermanifold. As a demonstration of this formula, we compute the renormalization group equations for two-fermion operators at the dimension-eight level in the Standard Model Effective Field Theory.
}

\allowdisplaybreaks

\begin{document} 
\maketitle
\flushbottom

%%%%%%%%%%%%%%%%%%%%%%%%%%%%%%%%%%%%%%%%%%%%%%%%%%%%%%%
\section{Introduction}\label{sec:Introduction}
%%%%%%%%%%%%%%%%%%%%%%%%%%%%%%%%%%%%%%%%%%%%%%%%%%%%%%%

The Standard Model Effective Field Theory (SMEFT) is endowed with a geometry in field space. This is well known in the scalar sector, where higher-dimensional Higgs operators that multiply the kinetic term can be reorganized into a metric on field space~\cite{Alonso:2015fsp,Alonso:2016oah}. However, this feature is not unique to scalars. Fermions and gauge bosons also enjoy a geometric description \cite{Helset:2020yio,Finn:2020nvn,Helset:2022tlf,Helset:2022pde,Assi:2023zid,Gattus:2023gep,Derda:2024jvo,Gattus:2024ird}. The underlying reason to expect such a description to exist comes from a rather mundane fact about any quantum field theory; the field variables can be redefined without changing the physics, since the quantum fields are integration variables in the path integral \cite{Chisholm:1961tha,Kamefuchi:1961sb,Politzer:1980me,tHooft:1973bhk,Arzt:1993gz,Cohen:2024fak}. Although the original applications of field-space geometry have a long history \cite{Volkov:1973vd}, recently these ideas have gone through a renaissance \cite{Alonso:2015fsp,Alonso:2016oah,Helset:2018fgq,Corbett:2019cwl,Helset:2020yio,Hays:2020scx,Corbett:2021eux,Cheung:2021yog,Cheung:2022vnd,Helset:2022tlf,Helset:2022pde,Assi:2023zid,Derda:2024jvo,Helset:2024vle,Cohen:2025dex,Cohen:2021ucp,Cohen:2022uuw,Craig:2023wni,Cohen:2023ekv,Cohen:2024bml,Li:2024ciy,Craig:2023hhp,Lee:2024xqa,Alminawi:2023qtf, Assi:2024zap,Assi:2025zmp,Finn:2019aip,Finn:2020nvn,Gattus:2023gep,Gattus:2024ird,Corbett:2023yhk,Martin:2023fad}. 

The field-space geometry can be used as a practical tool for computations. By manifesting the covariance under field redefinitions, many results are simplified when written in terms of geometric quantities such as the curvature in field space. For example, the tree-level scattering amplitude of two scalars and two fermions depends on the curvature of the scalar-fermion manifold \cite{Assi:2023zid}. Loop computations also benefit from a geometric treatment. The renormalization group equations (RGEs) of a general effective field theory (EFT) of scalars, fermions, and gauge bosons can be expressed in terms of geometric quantities, which compactly combine seemingly unrelated terms including higher-dimensional operators in the EFT.  This was used to compute the one-loop RGEs for the SMEFT to mass-dimension eight arising from boson and fermion loops \cite{Helset:2022pde,Assi:2023zid}. This result has also been extended to two loops for scalar EFTs \cite{Jenkins:2023bls,Jenkins:2023rtg}. 

To incorporate fermions cohesively into the geometric framework, we extend the field-space geometry to a supergeometry\footnote{
The supergeometry setup does not require the underlying theory to be supersymmetric. For example, we apply our technology to the generic SMEFT.
} \cite{Finn:2020nvn,Assi:2023zid}, where the field space becomes a supermanifold comprising both bosonic (commuting) and fermionic (anticommuting) coordinates. The supermanifold structure allows us to define supermetrics, superconnections, and supercurvature that unify the treatment of scalars, fermions, and gauge bosons within a single geometric framework. We will show how this approach is used to derive one-loop RGEs involving mixed boson-fermion graphs while preserving covariance under field redefinitions.

The SMEFT RGEs have also been computed using diagrammatic, functional, and on-shell methods. The complete dimension-six one-loop results were computed in refs.~\cite{Jenkins:2013zja,Jenkins:2013wua,Alonso:2013hga,Alonso:2014zka}. For the dimension-eight operators, the one-loop RGE results are incomplete. Currently, known results include the bosonic sector \cite{Helset:2022pde,Assi:2023zid,Chala:2021pll,DasBakshi:2022mwk,AccettulliHuber:2021uoa} and parts of the two-fermion sector \cite{Bakshi:2024wzz} and the four-fermion sector \cite{Boughezal:2024zqa}. In addition, the two-loop dimension-six RGE results for the bosonic sector were recently evaluated using functional methods \cite{Fuentes-Martin:2023ljp,Born:2024mgz} and geometric methods \cite{Jenkins:2023bls}. RGEs including lepton-number-violating operators were also computed at one-loop order in refs.~\cite{Antusch:2001ck,Liao:2016hru, Liao:2019tep, Davidson:2018zuo, Chala:2021juk, Zhang:2023kvw,Zhang:2023ndw,DasBakshi:2023htx} and at two-loop order in ref.~\cite{Ibarra:2024tpt}.
Generic one-loop RGE results for an arbitrary EFT up to dimension six were given in refs.~\cite{Fonseca:2025zjb,Misiak:2025xzq,Aebischer:2025zxg}. There is also an ongoing effort to extend the one-loop RGEs in the low-energy effective theory below the electroweak scale (LEFT) \cite{Jenkins:2017dyc} to two-loop order~\cite{Naterop:2024ydo,Aebischer:2025hsx}.
Automation of one- and two-loop RGE calculations for generic EFTs is currently underway in \texttt{Matchete} \cite{Fuentes-Martin:2022jrf} and \texttt{MatchMakerEFT} \cite{Carmona:2021xtq}. Finally, on-shell methods can give particular insights into the structure of the anomalous-dimension matrix \cite{Alonso:2014rga,Cheung:2015aba,Bern:2019wie,Bern:2020ikv,Chala:2023xjy,Chala:2023jyx,Craig:2019wmo,Bresciani:2023jsu,Baratella:2020lzz,Baratella:2020dvw,EliasMiro:2020tdv,AccettulliHuber:2021uoa,DelleRose:2022ygn}, which can be further constrained by flavor selection rules \cite{Machado:2022ozb}.

This paper is organized as follows. We introduce the Lagrangian and the geometric framework in \cref{sec:Formalism}, while in \cref{sec:Renormalization} we derive the renormalization formula for mixed boson-fermion graphs. In \cref{sec:Variations}, we compute the first and second variations of the action needed for the renormalization formula. The variations are explicitly covariant under field redefinitions, thereby guaranteeing that the final result shares this property. The application of this result to the SMEFT is discussed in \cref{sec:SMEFT}, before we conclude in \cref{sec:Conclusion}. The appendices include details on the field-space geometry (\cref{sec:Geometry}), a discussion of covariance (\cref{sec:Covariance}), the operator basis (\cref{sec:Operators}), and the renormalization group results for two-fermion dimension-eight operators in the SMEFT (\cref{sec:RGEresults}).

%%%%%%%%%%%%%%%%%%%%%%%%%%%%%%%%%%%%%%%%%%%%%%%%%%%%%%%
\section{Geometric Formalism}\label{sec:Formalism}
%%%%%%%%%%%%%%%%%%%%%%%%%%%%%%%%%%%%%%%%%%%%%%%%%%%%%%%

The effective field theory we consider takes the form
\begin{align}\label{eq:Lagr}
    \cL =& \frac{1}{2} h_{IJ}(\phi) (D_{\mu} \phi)^{I} (D^{\mu} \phi)^{J} - V(\phi) - \frac{1}{4} g_{AB}(\phi) F^{A}_{\mu\nu} F^{B\mu\nu} 
     \\
    &+ \frac{1}{2} i k_{\bar p r}(\phi) \left(\bar \psi^{\bar p} \overset{\leftrightarrow}{\slashed{D}} \psi^{r}\right) + i \omega_{\bar p r I}(\phi) (D_{\mu} \phi)^{I} \left(\bar \psi^{\bar p} \gamma^{\mu} \psi^{r}\right)  - \bar \psi^{\bar p}\left( \cM_{\bar p r}(\phi)-\sigma_{\mu\nu} \cT_{\bar p r}^{\mu\nu}(\phi,F) \right)\psi^{r}  \nonumber \,.
\end{align}
We use the indices $I,J,K,\dots$ for scalars, $A,B,C,\dots$ for gauge fields, and $p,\bar p,r,\bar r,\dots$ for fermions. All functions $h_{IJ}(\phi)$, $V(\phi)$, $g_{AB}(\phi)$, $k_{\bar p r}(\phi)$, $\omega_{\bar p r I}(\phi)$, and $\cM_{\bar p r}(\phi)$, %and $C_{\bar p r \bar s t}(\phi)$ 
are functions of the scalar field, following the approach in ref.~\cite{Helset:2020yio}, while $\cT^{\mu\nu}_{\bar p r}(\phi,F)=\cT_{\bar p r A}(\phi)F^{A\mu\nu}$ also depends on the field strength. 

To incorporate fermions into the geometric framework, we extend the field-space geometry to a \emph{supermanifold}, where the field space comprises both bosonic (commuting) and fermionic (anticommuting) coordinates~\cite{DeWitt:2012mdz,rogers2007supermanifolds}. On this supermanifold, scalar fields and gauge bosons are treated as coordinates on the bosonic submanifold, while fermionic fields are coordinates on the fermionic submanifold.

To establish some notation, consider the case of scalars and fermions.
We group the scalars and fermions into a multiplet to chart the manifold:
\begin{align}\label{eq:chart_scalarfermion}
    \Phi^{a} = 
    \begin{pmatrix}
        \phi^{I} \\
        \psi^{p} \\
        \bar \psi^{\bar p}
    \end{pmatrix} \,.
\end{align}
Here, the indices $a,b,c,\dots$ run over both bosonic and fermionic coordinates of the supermanifold. Specifically, $a$ can represent a scalar index $I$, a fermion index $p$, or a barred fermion index $\bar p$. The supermanifold indices carry \emph{Grassmann parity} $\epsilon_a$, which is $0$ for bosonic indices and $1$ for fermionic indices.

In constructing the geometric framework, we introduce a supermetric ${}_a\bar g_{b}(\Phi)$ on the supermanifold, which encodes the kinetic terms of the fields and part of their interactions. The placement of indices is a feature in supergeometry due to the anticommuting nature of fermionic coordinates. The left indices indicate that the index $a$ is associated with the left position in tensor products, which is important for tracking signs in supergeometry. For example, when contracting tensors, the order matters due to the anticommuting nature of fermionic coordinates:
\begin{align}
    (T \cdot V)^a = T^{ab} V_b = (-1)^{\epsilon_a \epsilon_b} V_b T^{ab} \,,
\end{align}
where $\epsilon_a$ and $\epsilon_b$ are the Grassmann parities of the indices.

Starting from the scalar-fermion part of the Lagrangian,
\begin{align}\label{eq:Lagrsc}
    \cL \supset \frac{1}{2} h_{IJ}(\phi) (\partial_{\mu} \phi)^{I} (\partial^{\mu} \phi)^{J}
    + \frac{1}{2} i k_{\bar p r}(\phi) (\bar \psi^{\bar p}  \overset{\leftrightarrow}{\slashed{\partial}} \psi^{r}) + i \omega_{\bar p r I}(\phi) (\partial_{\mu} \phi)^{I} \bar \psi^{\bar p} \gamma^{\mu} \psi^{r}  \,,
\end{align}
we find that the supermetric components are given by \cite{Assi:2023zid}
\begin{align}\label{eq:supermetric}
    {}_a\bar g_{b}(\Phi) 
     = \begin{pmatrix}
        h_{IJ} & (\bar{\psi} \omega^- )_{rI} & (\omega^+ \psi )_{\bar{r}I} \\
        -(\bar{\psi} \omega^- )_{pJ}  & 0 & k_{\bar r p} \\
       -(\omega^+ \psi )_{\bar{p}J} & - k_{\bar p r} & 0
    \end{pmatrix} \,,
\end{align}
with $(\bar \psi \omega^-)_{rI}= \bar \psi^{\bar p} \omega^-_{\bar p r I}$ and  $(\omega^+ \psi)_{\bar p J}= \omega^+_{\bar p r I} \psi^{r} $, and where we have defined
\begin{align}\label{eq:omegapm}
    \omega^{\pm}_{\bar p r I} = \omega_{\bar p r I} \pm \frac{1}{2} k_{\bar p r,I} \,,
\end{align}
with $k_{\bar p r,I} = \partial_{I} k_{\bar p r}$. The supermetric is supersymmetric\footnote{The term supersymmetry here refers to a property of the supermetric, specifically, that it is equal to its supertranspose. This does not mean the theory has to be supersymmetric in the field theory sense.} and %its components obey certain symmetry properties that depend on the Grassmann parity of the indices. Specifically, 
the following relations hold for shuffling indices~\cite{DeWitt:2012mdz}:
\begin{align}\label{eq:supermetric_symmetry}
\begin{aligned}
       \bar g_{ab} &= (-1)^{\epsilon_a} \, {}_a\bar g_{b} \,, \quad  {}_a\bar g_{b} = (-1)^{\epsilon_a + \epsilon_b + \epsilon_a \epsilon_b} \, {}_b\bar g_{a} \,, \quad  \bar g_{ab} = (-1)^{\epsilon_a \epsilon_b} \bar g_{ba} \,, \\
     \bar g^{ab} &= {}^a\bar g^b \,, \quad   {}^a\bar g^b = (-1)^{\epsilon_a \epsilon_b} \, {}^b\bar g^a \,, \quad  \bar g^{ab} = (-1)^{\epsilon_b \epsilon_a} \bar g^{ba} \,.
\end{aligned}
\end{align}
The placement of indices together with the Grassmann parity ensure the correct behavior under permutations, taking into account the signs introduced by anticommuting fermionic coordinates.

By incorporating fermions into the field-space geometry using supermanifolds, we can define geometric quantities that include both bosonic and fermionic fields. This unified framework preserves covariance under nonderivative field redefinitions for all types of fields and enables us to express physical quantities such as scattering amplitudes and renormalization group equations in terms of geometric invariants of the supermanifold.

Further details on the supermetric and additional properties such as incorporating gauge bosons are provided in \cref{sec:Geometry}, where we also discuss how the supermetric is used to define covariant derivatives, connections, and curvature tensors on the supermanifold.

%%%% CUT-OFF (new section?)
%%%%%%%%%%%%%%%%%%%%%%%%%%%%%%%%%%%%%%%%%%%%%%%%%%%%%%%
\section{Renormalization from mixed boson-fermion graphs}\label{sec:Renormalization}

Anomalous dimensions at one-loop order can be efficiently computed for general field theories. In ref.~\cite{tHooft:1973bhk}, the UV-divergences, and thereby the renormalization counterterms, for a renormalizable theory of scalars, fermions, and gauge bosons were extracted from a compact formula. This result was then extended to boson loops~\cite{Alonso:2016oah,Helset:2022pde} and fermion loops~\cite{Assi:2023zid} for effective field theories at one-loop order, and to scalar effective field theories at two-loop order \cite{Jenkins:2023rtg,Jenkins:2023bls}. Crucially, these results employed a geometric framework. Here, we complete this computation by including mixed boson-fermion graphs for effective field theories at one-loop order.  

To set the stage, consider the computation in the bosonic sector. We use the background field method to expand the action to second order in the fluctuation fields, $\phi \rightarrow \phi_{\textrm{B}} + \eta$. Following ref.~\cite{Helset:2022pde}, we group the scalars and gauge fields into multiplets, e.g., the fluctuation field is
\begin{align}
    \eta^{i} = 
    \begin{pmatrix}
        \eta^{I} \\
        \zeta^{A\mu_A}
    \end{pmatrix} \,,
\end{align}
and the index $i$ runs over both scalar and gauge field indices. After covariantly expanding the action to second order in the fluctuation fields, we obtain
\begin{align}
    \label{eq:Setaeta}
    \nabla_{\eta\eta} S = \frac{1}{2} \int \rd^{4}x \left\{  (\mathcal{D}_{\mu}\eta^{i}) \;_{i}g_{j}(\mathcal{D}^{\mu}\eta^{j}) + X_{ij} \eta^{i} \eta^{j} \right\} \,.
\end{align}
The covariant derivative depends on the scalar--gauge-boson geometry \cite{Helset:2022pde},
\begin{align} \label{eq:SCovDer}
    (\cD_{\mu}\eta)^{i} = (D_\mu \eta)^{i} + \Gamma^{i}_{jk} Z^{j}_{\mu}\eta^{k} \,, 
    \qquad
    Z^{i}_{\mu} = \begin{pmatrix}
        (D_{\mu}\phi)^{I} \\
        F^{A}_{\mu\nu_A}
    \end{pmatrix} \,,
\end{align}
where $D_\mu$ is the gauge covariant derivative.\footnote{After introducing the fermions into the geometry, the new term in \cref{eq:SCovDerfermions} will appear in the definition for the covariant derivative of the scalar fluctuation.}
Remarkably, the renormalization formula for boson loops for effective field theories is simply the covariantized version of the original formula derived for renormalizable theories \cite{tHooft:1973bhk},
\begin{align} 
\label{eq:OneLoopDivBoson}
	\Delta S_{\rm boson} =& \frac{1}{32 \pi^2 \epsilon}   \int \rd^{4}x \left\{
    \frac{1}{12} {\rm Tr}\left[Y_{\mu\nu} Y^{\mu\nu}\right] + \frac{1}{2} {\rm Tr}\left[X^2\right] \right\} \,,
\end{align}
where $\left[Y_{\mu\nu}\right]^{i}_{\;\;j} = \left[\cD_{\mu}, \cD_{\nu}\right]^{i}_{\;\;j}$. 

For the fermions, the situation is similar. We use the background field method, $\psi \rightarrow \psi_{\textrm{B}} + \chi$, where $\chi$ is the fermion fluctuation field and $\psi_{\textrm{B}}$ is the background field. The covariant second variation of the action for the fermions is
\begin{align}
    \label{eq:Schichi}
    \nabla_{\bar \chi \chi}S = \int \rd^{4}x \left\{ \frac{1}{2}i k_{\bar pr} \left( \bar \chi^{\bar p} \overset{\leftrightarrow}{\slashed \cD} \chi^{r}\right) - \bar \chi^{\bar p} \cM_{\bar p r} \chi^{r} + \bar \chi^{\bar p} \sigma_{\mu\nu}\cT^{\mu\nu}_{\bar p r} \chi^{r} \right\} \,.
\end{align}
The covariant derivative acts on the fermion fields as
\begin{align}\label{eq:defGeomCovD}\begin{aligned}  
    \cD_{\mu}\psi^{r} &= D_{\mu} \psi^{r} + \bar \Gamma^{r}_{Is} (D_{\mu}\phi)^{I} \psi^{s} \,, \\
    \cD_{\mu}\bar \psi_{r} &= D_{\mu} \bar \psi_{r} - \bar \Gamma^{s}_{Ir} (D_{\mu}\phi)^{I} \bar\psi_{s} \,. \\
\end{aligned}\end{align}
The mixed scalar-fermion Christoffel symbols are defined in \cref{sec:Geometry}.

Although the kinetic and mass terms in a renormalizable theory were taken into account in the original divergence formula, the dipole term was not. The full result is \cite{Assi:2023zid}
\begin{align} 
    \label{eq:OneLoopDivFermion}
	\Delta S_{\rm fermion} =& \frac{1}{32 \pi^2 \epsilon}   \int \rd^4x \ \bigg\{ 
  \frac{1}{3} {\rm Tr} \left[ \cY_{\mu\nu} \cY^{\mu\nu} \right] + {\rm Tr} \left[ (\cD_{\mu} \cM) (\cD^{\mu} \cM) - (\cM\cM)^2 \right] \\
 & \qquad \qquad\qquad -\frac{16}{3} {\rm Tr}[(\cD_\mu \mathcal T^{\mu \alpha})(\cD_\nu \mathcal T^{\nu \alpha})  - (\mathcal T^{\mu\nu} \mathcal T^{\alpha \beta})^2 ] \nonumber \\
& \qquad \qquad\qquad -4 i {\rm Tr}[ \cY_{\mu\nu} (\mathcal M \mathcal{T^{\mu\nu}}+ \mathcal T^{\mu\nu}\mathcal M)] - 8 {\rm Tr}[ (\mathcal M \mathcal T^{\mu\nu})^2] \bigg\} \,, \nonumber
\end{align}
where $\left[\cY_{\mu\nu}\right]^{p}_{\;\;r} = \left[\cD_{\mu}, \cD_{\nu}\right]^{p}_{\;\;r}$. Note that $\mathcal Y_{\mu\nu}$ carry fermionic indices, in contrast to the bosonic indices of $Y_{\mu\nu}$ defined above.

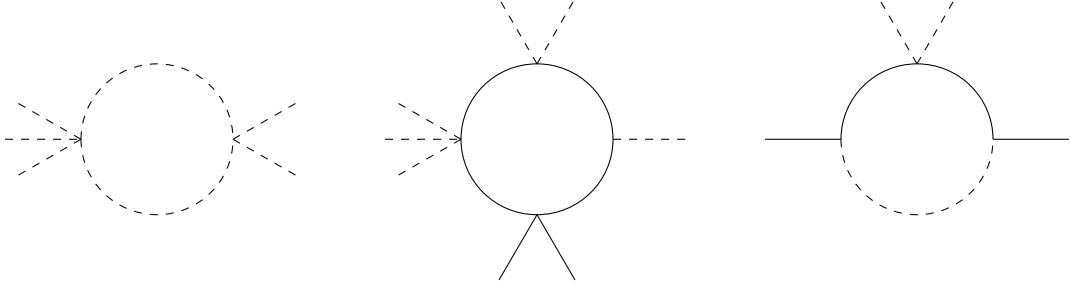
\begin{figure}
\begin{center}
\begin{tikzpicture}
% Define the radius of the loop
  \def\radius{1cm}

  % Starting point for the left solid line
  \draw[dashed] (-2,0) -- (-\radius,0);
  \draw[dashed] (-1,0) -- +(30:-1); % first dashed line
  \draw[dashed] (-1,0) -- +(-30:-1); % second dashed line
  
  % Solid half of the loop (right half)
  \draw[dashed] (-\radius,0) arc (180:0:\radius);

  % Dashed half of the loop (left half)
  \draw[dashed] (-\radius,0) arc (-180:0:\radius);

  % Ending point for the right solid line
  %\draw[dashed] (\radius,0) -- (2,0);
  \draw[dashed] (1,0) -- +(30:1); % first dashed line
  \draw[dashed] (1,0) -- +(-30:1); % second dashed line
    % Two dashed lines emanating from the same point on the right leg

  %\draw[dashed] (-0,-1) -- +(60:-1); % first dashed line
  %\draw[dashed] (-0,-1) -- +(120:-1); % second dashed line
  %\draw[dashed] (0,1) -- +(-60:-1); % first dashed line
  %\draw[dashed] (0,1) -- +(-120:-1); % second dashed line
\begin{scope}[shift={(5,0)}]
  % Define the radius of the loop
  \def\radius{1cm}

  % Starting point for the left solid line
  \draw[dashed] (-2,0) -- (-\radius,0);
   \draw[dashed] (-1,0) -- +(30:-1); % first dashed line
  \draw[dashed] (-1,0) -- +(-30:-1); % second dashed line
  
  % Solid half of the loop (right half)
  \draw (-\radius,0) arc (180:0:\radius);

  % Dashed half of the loop (left half)
  \draw (-\radius,0) arc (-180:0:\radius);

  % Ending point for the right solid line
  \draw[dashed] (\radius,0) -- (2,0);
    % Two dashed lines emanating from the same point on the right leg

  \draw (-0,-1) -- +(60:-1); % first dashed line
  \draw (-0,-1) -- +(120:-1); % second dashed line
  \draw[dashed] (0,1) -- +(-60:-1); % first dashed line
  \draw[dashed] (0,1) -- +(-120:-1); % second dashed line
\end{scope}
\begin{scope}[shift={(10,0)}]
  % Define the radius of the loop
  \def\radius{1cm}

  % Starting point for the left solid line
  \draw (-2,0) -- (-\radius,0);
  
  % Solid half of the loop (right half)
  \draw (-\radius,0) arc (180:0:\radius);

  % Dashed half of the loop (left half)
  \draw[dashed] (-\radius,0) arc (-180:0:\radius);

  % Ending point for the right solid line
  \draw (\radius,0) -- (2,0);
    % Two dashed lines emanating from the same point on the right leg
  \draw[dashed] (0,1) -- +(-60:-1); % first dashed line
  \draw[dashed] (0,1) -- +(-120:-1); % second dashed line
\end{scope}
\end{tikzpicture}
\end{center}
\caption{A sample of one-loop bosonic, fermionic, and mixed corrections to the action. The solid line is the fermion propagator, and the dashed lines represent scalar or gauge fields. Each vertex can consist of any number, including zero, of background fields, represented by the external lines. \label{fig:loop}}
\end{figure}

The last contribution to the RGEs is from mixed boson-fermion graphs. In order to compute them, we need the mixed second variation of the action. Again, keeping all expressions covariant, we find that
\begin{align}
    \label{eq:Setachi}
    \nabla_{\eta\chi} S 
    &= \int d^4 x \left[
    \left( \bar \chi^{\bar p} N_{\bar p i} + \bar N_{r i} \chi^{r} \right) \eta^{i}
    +
    \left( \bar \chi^{\bar p}\gamma^{\mu} Q_{\bar p i} + \bar Q_{r i} \gamma^{\mu} \chi^{r} \right) (\cD_\mu \eta^{i})
    \right] \,.
\end{align}
From this we can compute the one-loop UV divergence. We split the result into two parts, one with two insertions of the mixed vertices from \cref{eq:Setachi}, and one with four such insertions. The result for two insertions is 
\begin{align} 
\label{eq:OneLoopDivMix2}
	\Delta S_{\rm mix}^{(2)} =& \frac{1}{32 \pi^2 \epsilon}   \int \rd^4x \ \bigg\{
 {\rm Tr}[ \bar N (i \slashed \cD + 2 \cM) N ]
 - 2{\rm Tr}[ i\bar Q N X + {\rm h.c.}]
  \\
 & \qquad \qquad\qquad + {\rm Tr}[ i\bar Q (i \overset{\leftarrow}{\slashed \cD} + \cM - \sigma_{\alpha\beta} \cT^{\alpha\beta} )(i \slashed \cD + 2 \cM) N + {\rm h.c.}]
 \nonumber \\
 & \qquad \qquad\qquad + {\rm Tr}[ \bar Q (i \slashed \cD - \cM + \sigma_{\alpha\beta} \cT^{\alpha\beta} )Q X + {\rm h.c.}]
  \nonumber \\
 & \qquad \qquad\qquad - {\rm Tr}[ \bar Q (i \overset{\leftarrow}{\slashed \cD} + \cM - \sigma_{\alpha\beta} \cT^{\alpha\beta} )(i \slashed \cD + 2 \cM)(i \slashed \cD - \cM + \sigma_{\gamma\delta} \cT^{\gamma\delta} )Q ] \,.
 \nonumber 
\end{align}
For four insertions, we get
\begin{align} 
\label{eq:OneLoopDivMix4}
	\Delta S_{\rm mix}^{(4)} = & \frac{1}{32 \pi^2 \epsilon}   \int \rd^4x \ \bigg\{  \frac{1}{2}{\rm Tr}[  (\bar Q \gamma_\mu Q ) (\bar N \gamma^\mu N ) ]
  + 2{\rm Tr}[(\bar Q  N ) (\bar N  Q )]
-  {\rm Tr}[  (\bar Q N ) (\bar Q  N )  + {\rm h.c.}]
  \nonumber \\
 & \qquad \qquad\qquad -2 {\rm Tr}[ i\bar Q (i \slashed \cD - \cM+\sigma_{\alpha\beta}\cT^{\alpha\beta}) Q)(\bar{N}Q) + {\rm h.c.}] 
   \\
 & \qquad \qquad\qquad + \frac{1}{2}{\rm Tr}[  i(\bar Q \gamma^\mu Q )(\bar N\gamma_\mu(i\overset{\leftarrow}{\slashed\cD}+\cM-\sigma_{\alpha\beta}\cT^{\alpha\beta})  Q ) +\rm{h.c.} ] \,
  \nonumber \\
 & \qquad \qquad\qquad - \frac{1}{2}{\rm Tr}[  (\bar Q\gamma^\mu Q )(\bar Q \gamma_\mu Q )X +\rm{h.c.}]
   \nonumber \\
 & \qquad \qquad\qquad - {\rm Tr}[ \bar Q (i \overset{\leftarrow}{\slashed \cD} + \cM - \sigma_{\alpha\beta} \cT^{\alpha\beta} )Q\bar Q(i \slashed \cD - \cM + \sigma_{\gamma\delta} \cT^{\gamma\delta} )Q]
    \nonumber \\
 & \qquad \qquad\qquad - {\rm Tr}[ \bar Q (i \overset{\leftarrow}{\slashed \cD} + \cM - \sigma_{\alpha\beta} \cT^{\alpha\beta} )\gamma^{\mu}(i \slashed \cD - \cM + \sigma_{\gamma\delta} \cT^{\gamma\delta} )Q\bar Q\gamma_{\mu}Q]
\bigg\} \,. \nonumber
\end{align}
There are also contributions from six or eight insertions, but we do not compute them because they will renormalize six- and eight-fermion operators, showing up starting at dimension 9, with little phenomenological relevance.
In \cref{fig:loop}, the structure of the boson, fermion, and mixed boson-fermion graphs are shown. Each vertex can have any number of bosonic and fermionic background fields attached.
%%%%%%%%%%%%%%%%%%%%%%%%%%%%%%%%%
%%%%%%%%%%%%%%%%%%%%%%%%%%%%%%%%%

%%%%%%%%%%%%%%%%%%%%%%%%%%%%%%%%%%%%%%%%%%%%%%%%%%%%%%%
\section{Variations of the action}\label{sec:Variations}
%%%%%%%%%%%%%%%%%%%%%%%%%%%%%%%%%%%%%%%%%%%%%%%%%%%%%%%
%

We need the second variations of the action for the one-loop renormalization computation. To keep covariance at each step in the computation, we employ the background-field expansion in geodesic coordinates \cite{Honerkamp:1971sh,Alonso:2015fsp,Alonso:2016oah}. For a scalar, this is
\begin{align}
    \phi^{I} = \phi^{I}_{B} + \eta^{I} - \frac{1}{2}\Gamma^{I}_{JK}\eta^{I}\eta^{J} + \mathcal{O}(\eta^3) \,.
\end{align}
In the full theory, we also have gauge bosons and fermions. To preserve covariance, we group the fields into multiplets, where the combined fluctuation field is
\begin{align}
    \eta^{a} = 
    \begin{pmatrix}
        \eta^{I} \\
        \zeta^{A\mu_A} \\
        \chi^{p} \\
        \bar \chi^{\bar p}
    \end{pmatrix} \,.
\end{align}
The geodesic expansion now reads \cite{Helset:2022pde}
\begin{align}
    \Phi^{a} = \Phi^{a}_{B} + \eta^{a} - \frac{1}{2} \bar\Gamma^{a}_{bc} \eta^{b}\eta^{c} + \mathcal{O}(\eta^3) \,.
\end{align}
The Christoffel symbols of the supermanifold are given in \cref{sec:Geometry}.

\subsection{First variation of the action}

To start, let us consider the first variation of the action. 
The bosonic computation, captured by $\delta_{\eta}S_{\rm boson}$ and $\delta_{\zeta}S_{\rm boson}$, was previously derived in ref.~\cite{Helset:2022pde}. For the fermion operators, we find that
\begin{align}
    \label{eq:Seta}
    \delta_{\eta} S =& \delta_{\eta} S_{\rm boson} + 
    \int d^{4}x \bigg\{ - \bar \psi^{\bar p} \cM_{\bar p r} \psi^{r} 
    %\nonumber \\ & \qquad\qquad\qquad\qquad
    + \bar \psi^{\bar p} \sigma_{\mu\nu} F^{A}_{\mu\nu} \cT_{\bar p r A} \psi^{r} 
    %+ C_{\bar p r \bar s t} \left( \bar \psi^{\bar p} \gamma^{\mu} \psi^{r} \right)\left( \bar \psi^{\bar s} \gamma_{\mu} \psi^{t} \right)
    \bigg\}\,_{,I}\, \eta^{I}
    \nonumber \\ & \qquad\quad\;\;
    +\int d^{4}x \bigg\{
    i \left((\bar \nabla_{I}\bar\psi_{r}) \gamma^{\mu} (\cD_{\mu} \psi^{r}) \right)
    - i \left((\cD_{\mu} \bar\psi_{r}) \gamma^{\mu} (\bar \nabla_{I}\psi^{r}) \right)
    \nonumber \\ & \qquad\qquad\qquad\qquad
    + i \bar R_{\bar p r I J} (D_{\mu}\phi)^{J} \left( \bar \psi^{\bar p} \gamma^{\mu} \psi^{r} \right)
    \bigg\} \eta^I
    \,, \\
    \label{eq:Szeta}
    \delta_{\zeta} S =& \delta_{\zeta} S_{\rm boson} 
    +\int d^{4}x \bigg\{
    i \bar \psi^{\bar p} (t_{\bar p A,r} + \omega_{\bar prI}t^{I}_{A} ) \gamma_{\mu}\psi^{r}
    \bigg\} \zeta^{A\mu}
    \nonumber \\ & \qquad\quad\;\;
    +\int d^{4}x \bigg\{
    -2 \bar \psi^{\bar p} \sigma_{\mu\nu} \cT_{\bar p r A} \psi^{r}
    \bigg\} (D_{\nu}\zeta^{A\mu}) \,, \\
    \label{eq:Schi}
    \delta_{\chi} S =& 
    \int d^{4}x \bigg\{ - i \left((\cD_{\mu} \bar\psi_{r}) \gamma^{\mu} \psi^{r} \right) - \bar \psi^{\bar p} \cM_{\bar p r} \psi^{r} 
    + \bar \psi^{\bar p} \sigma_{\mu\nu} F^{A}_{\mu\nu} \cT_{\bar p r A} \psi^{r} \bigg\}\,_{,s}\,\chi^{s}  + \textrm{h.c.} 
\end{align}
where $X_{,s}$ is understood as $X\frac{\overleftarrow\partial}{\partial \psi^s }$.
Here we have used that 
\begin{align}
    \label{eq:nablaPsi}
    \bar \nabla_{I}\psi^{r} &= \partial_{I} \psi^{r} + \bar \Gamma^{r}_{Is} \psi^{s} = + k^{r \bar t} \omega^{+}_{\bar t s I} \psi^{s} \,, \\
    \label{eq:nablaPsiBar}
    \bar \nabla_{I}\bar \psi_{r} &= \partial_{I} \bar \psi_{r} - \bar \Gamma^{s}_{Ir} \bar \psi_{s} = -  \omega^{-}_{\bar t r I} \psi^{\bar t} \,,
\end{align}
where in the last equalities in \cref{eq:nablaPsi,eq:nablaPsiBar} we used $\partial_{I} \psi = 0$.

The first variation of the action is organized in such a way as to simplify the transformation properties under field redefinitions.
We will explicitly show that the first variations are covariant in \cref{sec:Covariance}.

\subsection{Second variation of the action}

The second variations of the action are
\begin{align}
    \label{eq:variationEtaEta}
    \nabla_{\eta\eta} S_{\rm fermion} 
    =& \frac{1}{2} \int d^{4}x
    \bigg\{ - \bar \psi^{\bar p} \cM_{\bar p r} \psi^{r} 
    %\nonumber \\ & \qquad\qquad\qquad\qquad
    + \bar \psi^{\bar p} \sigma_{\mu\nu} F^{A}_{\mu\nu} \cT_{\bar p r A} \psi^{r} 
    \bigg\}_{;IJ} \eta^{I}\eta^{J} \nonumber \\ 
    +&\frac{1}{2}\int d^{4}x \bigg\{
    i \left((\bar \nabla_{I}\bar \nabla_{J}\bar\psi_{r}) \gamma^{\mu} (\cD_{\mu} \psi^{r}) \right)
    - i \left((\cD_{\mu} \bar\psi_{r}) \gamma^{\mu} (\bar \nabla_{I}\bar \nabla_{J}\psi^{r}) \right)
    \nonumber \\ &  \qquad\quad
    + i \left((\bar \nabla_{I}\bar\psi_{r}) \gamma^{\mu} (\cD_{\mu} \bar \nabla_{J}\psi^{r}) \right)
    - i \left((\cD_{\mu} \bar \nabla_{J}\bar\psi_{r}) \gamma^{\mu} (\bar \nabla_{I}\psi^{r}) \right)
    \nonumber \\ &  \qquad\qquad\qquad\qquad
    + i (\bar \nabla_{J}\bar R_{\bar p r I K}) (D_{\mu}\phi)^{K} \left( \bar \psi^{\bar p} \gamma^{\mu} \psi^{r} \right)
    \bigg\}\; \eta^I \eta^{J}
     \\
    &+ \int d^{4}x \left\{ i \frac{1}{2} \left[ \nabla_{I} \omega_{\bar prJ} - \nabla_{J} \omega_{\bar prI} \right] (\bar \psi^{\bar p} \gamma_{\mu} \psi^{r}) \right\} \eta^{I} (D_{\mu} \eta^{J} + \Gamma^{J}_{KL} (D_\mu\phi)^{K} \eta^{L})  \,,
    \nonumber \\
    \label{eq:variationEtaZeta}
\nabla_{\eta\zeta} S_{\rm fermion} 
    =&
    \int d^{4}x \left\{ 
    i \left[ (\delta^{B}_{A} \partial_{I} - \Gamma^{B}_{IA}) \left( t_{\bar p B,r} + \omega_{\bar p r J} t^{J}_{B} \right) \right] (\bar \psi^{\bar p} \gamma_{\mu} \psi^{r})
    \right\} \eta^{I}\zeta^{A\mu}
    \nonumber \\ &+ 2 
    \int d^{4}x  \,(\bar \psi^{\bar p} \sigma^{\mu\nu}  \psi^{r}) \left[  \cT_{\bar p r A,I} \eta^{I} D_\mu \zeta^A_\nu - \cT_{\bar p r A} D_{\mu}(\Gamma^{A}_{BI} \zeta^B_\nu \eta^{I}) \right] \,,
    \\ 
    \label{eq:variationEtaChi}
    \nabla_{\eta\chi} S_{\rm fermion} 
    =&
    \int d^4x \bigg\{ i \bar R_{\bar pr IJ}(D_{\mu}\phi)^{J}(\bar \psi^{\bar p}\gamma^{\mu}\chi^{r}) - i \left((\cD_{\mu}\bar\nabla_{I}\bar\psi_{r})\gamma^{\mu}\chi^{r} \right)
    \bigg\} \eta^{I}
    \nonumber \\ &+
    \int d^4x \bigg\{ \left[
    - (\bar \psi^{\bar p}\cM_{\bar pr}\psi^r) 
    + \bar \psi^{\bar p} \sigma_{\mu\nu} F^{A}_{\mu\nu} \cT_{\bar p r A} \psi^{r} 
    \right]_{;Is}\bigg\} \eta^{I}\chi^{s}
    \nonumber \\
    &+ \int d^{4}x \left\{ -i \left((\bar\nabla_I \bar \psi_{r}) \gamma_{\mu} \chi^r \right)  \right\} (D_{\mu} \eta^{I} + \Gamma^{I}_{JK}(D_\mu\phi)^{J}\eta^{K}) + \textrm{h.c.}\,,
    \\
\label{eq:variationZetaChi}
    \nabla_{\zeta\chi} S_{\rm fermion} 
    =&
    \int d^{4}x \left\{ i \left( t_{\bar p A,r} + \omega_{\bar p r I} t^{I}_{A} \right) (\bar \psi^{\bar p} \gamma_{\mu} \;\cdot\;) \right\} \chi^{r} \zeta^{A\mu}
    \nonumber \\ &
    +\int d^{4}x \bigg\{
    -2 \bar \psi^{\bar p} \sigma_{\mu\nu} \cT_{\bar p r A} \chi^{r}
    \bigg\} (D_{\nu}\zeta^{A\mu}) + \textrm{h.c.}\,, \\
    \label{eq:variationChiChi}
    \nabla_{\bar\chi \chi} S_{\rm fermion} 
    =&
    \int d^{4}x \left\{ 
    \frac{1}{2} k_{\bar pr} i \left( \bar \chi^{\bar p} \overset{\leftrightarrow}{\slashed \cD} \chi^{r} \right) 
    - \cM_{\bar pr} \left( \bar \chi^{\bar p} \chi^{r} \right)
    + \bar \chi^{\bar p} \sigma_{\mu\nu} F^{A}_{\mu\nu} \cT_{\bar p r A} \chi^{r}%\\ &\qquad\qquad \left.
    \right\} \,,\\
    \label{eq:variationZetaZeta}
    \nabla_{\zeta \zeta} S_{\rm fermion} 
    =&
    \left\{\frac12\frac{\delta_\eta S}{\delta \eta^I} \Gamma^I_{AB} \eta^{\mu\nu} 
    - \int d^4x \,\mathcal T_{\bar p r C} f^C_{~AB} (\bar \psi^{\bar p } \sigma^{\mu\nu} \psi^r) \right\} \zeta_\mu^A \zeta_\nu^B
    \,.
\end{align}
Here, we use the covariant second variation of the actions, $X_{;ij} = X \overset{\leftarrow}{\bar\nabla_{i}}\overset{\leftarrow}{\bar\nabla_{j}}$.
These expressions should be combined with their bosonic counterparts \cite{Helset:2022pde} to obtain the full second variation of the action.

Next, we need to rewrite these expressions to the forms in \cref{eq:Setaeta,eq:Schichi,eq:Setachi}. For instance, the scalar second variation of the action contains the term
\begin{align}
    \label{eq:SetaetaCrossTerm}
    \nabla_{\eta\eta}S_{\rm fermion} \supset \int d^{4}x \left\{ i \frac{1}{2} \left[ \nabla_{I} \omega_{\bar prJ} - \nabla_{J} \omega_{\bar prI} \right] (\bar \psi^{\bar p} \gamma_{\mu} \psi^{r}) \right\} \eta^{I} (D_{\mu} \eta^{J} + \Gamma^{J}_{KL} (D_\mu\phi)^{K} \eta^{L}) \,.
\end{align}
To match the form in \cref{eq:Setaeta}, we add a term to the covariant derivative of a scalar fluctuation eq.~\eqref{eq:SCovDer}.
\begin{align} \label{eq:SCovDerfermions}
    \cD_\mu \eta^{I} \supset +  \frac{i}{2}h^{IL}\left[ \nabla_{K}\omega_{\bar pr L} - \nabla_{L}\omega_{\bar pr K} \right](\bar\psi^{\bar p}\gamma_{\mu}\psi^{r}) \eta^{K} \,.
\end{align}
With this definition, the term in \cref{eq:SetaetaCrossTerm} can be absorbed into the scalar kinetic term at the expense of shifting the mass term $X_{ij}\eta^i\eta^j$ by a four-fermion operator. Note that the same definition for the covariant derivative must be used in the mixed scalar-fermion variations. 

A similar analysis applies to the term in the mixed scalar-gauge variation,
\begin{align}
    \nabla_{\eta\zeta}S \supset \int d^{4}x \left\{ 
    2 \bar \psi^{\bar p} \sigma_{\mu\nu} \bar \psi^{r}
    \right\}  \left[  \cT_{\bar p r A,I} \eta^{I} D_\mu \zeta^{A\nu}  \right] \,.
\end{align}
By redefining the mixed $\eta$-$\zeta$ terms in the covariant derivative on bosons, we can reabsorb this term in the first part of \cref{eq:Setaeta}. Also, in the mixed gauge-fermion variation, we encounter
\begin{align}
    \nabla_{\zeta\chi}S \supset \int d^{4}x \bigg\{
    -2 \bar \psi^{\bar p} \sigma_{\mu\nu} \cT_{\bar p r A} \chi^{r}
    \bigg\} (D_{\nu}\zeta^{A\mu}) \,.
\end{align}
With a judicious gauge-fixing choice, where the covariant gauge fixing in ref.~\cite{Helset:2018fgq} is extended to account for the dipole operator, we can massage this term into the form of \cref{eq:Setachi}.

Lastly, four-fermion operators can also be included in our analysis. 
Adding to the Lagrangian in \cref{eq:Lagr} a term
\begin{equation}
    \mathcal L \supset  C_{\bar p r \bar s t}(\phi) \left( \bar \psi^{\bar p} \gamma^{\mu} \psi^{r} \right)\left( \bar \psi^{\bar s} \gamma_{\mu} \psi^{t} \right) \,,
\end{equation}
modifies the definition of the supermetric in eq.~\eqref{eq:supermetric}.
As discussed in ref.~\cite{Assi:2023zid}, by exploiting Fierz identities,\footnote{Four-dimensional operations preserve the divergent structure at one-loop order.} we can place the fermion fluctuation fields in the same bilinear, and then absorb these terms in the mass, dipole, and kinetic terms in the fermion second variation of the action. 

Equipped with these expressions for the second variations of the action, we can now compute the one-loop UV divergences in the Standard Model Effective Field Theory.

%%%%%%%%%%%%%%%%%%%%%%%%%%%%%%%%%%%%%%%%%%%%%%%%%%%%%%%
\section{Application to the Standard Model Effective Field Theory}\label{sec:SMEFT}
%%%%%%%%%%%%%%%%%%%%%%%%%%%%%%%%%%%%%%%%%%%%%%%%%%%%%%%

The general results discussed above apply to the SMEFT. To fix the notation, we summarize the salient features of the geometric description for the SMEFT, which follows closely refs.~\cite{Helset:2020yio,Helset:2022pde,Assi:2023zid}. First, we group all the fields into multiplets. The complex Higgs doublet is expressed as a real multiplet
\begin{align}
    H = \frac{1}{\sqrt{2}} 
    \begin{pmatrix}
        \phi^2 + i \phi^1 \\
        \phi^4 - i \phi^3
    \end{pmatrix} 
    \qquad \Rightarrow \qquad
    \phi^{I} = \begin{pmatrix}
        \phi^1 \\
        \phi^2 \\
        \phi^3 \\
        \phi^4
    \end{pmatrix} 
    \,.
\end{align}
The fermion fields are grouped as
\begin{align}
    \label{eq:fermionMultiplet}
    \psi^{p} = 
    \begin{pmatrix}
        \ell^{p}_{L} \\
        q^{p}_{L} \\
        e^{p}_{R} \\
        u^{p}_{R} \\
        d^{p}_{R}
    \end{pmatrix} \,,
\end{align}
where $p$ runs over all type, flavor, and gauge-group indices. Periodically, we will abuse the notation by letting $p$ refer to all fermion indices, or just over the fermion gauge-group and flavor indices.\footnote{We started this practice already in \cref{eq:fermionMultiplet}.} The appropriate interpretation is understood from the context. Lastly, the gauge bosons from the gauge group $SU(3)_c \otimes SU(2)_L \otimes U(1)_{Y}$ are grouped as
\begin{align}
    A^{B}_{\mu} = \begin{pmatrix}
        G^{\mathscr{B}}_{\mu} \\
        W^{b}_{\mu} \\
        B_{\mu}
    \end{pmatrix} \,.
\end{align}
where $\mathscr{B}=1, \dots,8$ and $b=1,2,3$.
Next, we can specify the various functions in \cref{eq:Lagr} for the SMEFT. The scalar metric $h_{IJ}$, the gauge metric $g_{AB}$, and the potential $V$ are explicitly given in ref.~\cite{Helset:2022pde}. The fermion mass $\cM$, dipole $\cT$, and the fermion metric $k$ are found in ref.~\cite{Assi:2023zid}.

In addition, we need the boson-fermion vertices in \cref{eq:Setachi}. For our purposes, we will compute the linear-in-dimension-8 contributions to the RGEs, but ignore the dipole operators. The relevant parts of the boson-fermion vertices are
\begin{align}
     N_{\bar{p}I} \supset {}& \big(i\omega_{\bar{p}rJ,I}( D^\mu\phi^J) \gamma_\mu - \cM_{\bar{p}r,I}\big)\,\psi^r, \\
     N_{\bar{p}A\mu_A} \supset {}& i\big(t_{\bar{p}A,r}+\omega_{\bar{p}rI}t^I_A\big)\gamma_{\mu_A}\,\psi^r, \\
    Q_{\bar{p}I} \supset {}& i\omega_{\bar{p}rI}\psi^r \,.
\end{align}
The conjugate operators, \(\bar{N}_{ri}\) and \(\bar{Q}_{ri}\), are defined analogously. These expressions can be extracted from the second variations of the action in \cref{eq:variationEtaChi,eq:variationZetaChi}.

The last piece we need is $\omega_{\bar p r I}$. This term takes the form
\begin{align}
    \label{eq:omega}
    \omega_{\bar p r I} = \begin{pmatrix}
        \omega_{L,\bar p r I} & 0 \\
        0 & \omega_{R,\bar pr I}
    \end{pmatrix} \,.
\end{align}
The relevant operators are the $\psi^2 H^2 D$ and $\psi^2 H^4 D$ operators in \cref{dim:6_2fermion,dim:8_fermion}. For the left-handed fermions at dimension 8, we have
\begin{align}
\begin{aligned}
    \omega_{L,\bar p r I} \supset 
   & +\mathbb{I} \,\phi^2(\phi T_R^3)_{I} \coef{8}{\underset{\bar pr}{\ell^2H^4 D}}{(1)} 
    +4\, \sigma^a (\phi T_L^a T_R^3 \phi)(\phi T_R^3)_I  \coef{8}{\underset{\bar pr}{\ell^2H^4 D}}{(2)}
    \\
    &+ \sigma^a\,  \phi^2 (\phi T_L^a)_{I} \coef{8}{\underset{\bar pr}{\ell^2H^4 D}}{(3)}
    -4\, \epsilon_{abc} \sigma^a (\phi T_L^b T_R^3 \phi)(\phi T_L^c)_I \coef{8}{\underset{\bar pr}{\ell^2H^4 D}}{(4)}
  \,,
  \end{aligned}
\end{align}
where the suppressed indices on the identity matrix $\mathbb{I}$ and the Pauli matrices $\sigma^a$ correspond to the $SU(2)_L$ indices of the fermions. The matrices $T_L^a$ and $T_R^a$ with $a=1,2,3$ are given in appendix A of ref.~\cite{Helset:2022pde}. The contributions of the quark operators are analogous to the lepton operators. In contrast, for the right-handed fermions, the function takes the form
\begin{align}
    \omega_{R,\bar p r I} \supset 
    %+ \;i(\phi\gamma_4)_{I} \coef{6}{\underset{\bar pr}{e^2H^2 D}}{(1)} \,.
    + \phi^2 \left(\phi T_R^3\right)_I \coef{8}{\underset{\bar pr}{e^2H^4 D}}{} \,,
\end{align}
with analogous expressions for $u$ and $d$.
There is also an off-diagonal contribution to the right-handed part
\begin{align}
    \omega_{R,\bar p r I} \supset 
    %+ 
    -\frac12 \phi^2 \left(\phi (T_R^1- i T_R^2)\right)_I \coef{8}{\underset{\bar pr}{udH^4 D}}{} + \textrm{h.c.}
\end{align}
The chiral structure of $\omega$ in \cref{eq:omega} will be used heavily throughout the computation for terms such as $\omega_{\bar p r I} (\bar \psi^{\bar p} \psi^{r}) = 0$. The chirality of the fermion fields demands that one left-handed and one right-handed fermion is combined to form a Lorentz-invariant combination. The diagonal structure of \cref{eq:omega} then sets the term to zero. This feature is further discussed around \cref{eq:chiralVanishIdentity}.

We will compute the RGEs for some two-fermion operators with mass-dimension eight in the SMEFT, including effects from the bosonic operators and the $\psi^2 H^{2n+1}$ and $\psi^2 H^{2n} D$ operators in the Lagrangian in \cref{eq:Lagr}. 
The operator basis for the SMEFT at dimension six and dimension eight that adheres to our starting Lagrangian in \cref{eq:Lagr} is given in \cref{dim:6,dim:6_2fermion,dim:8,dim:8_fermion}%dim:6_4fermion,
. First, we pack the operators into the functions of the Lagrangian, such as the metric, potential, etc. Then we use these explicit expressions to compute the boson loop in \cref{eq:OneLoopDivBoson}, the fermion loop in \cref{eq:OneLoopDivFermion}, and the mixed boson-fermion loop in \cref{eq:OneLoopDivMix2}. The output of this computation is typically not in our operator basis, but rather in the off-shell Green's basis \cite{Chala:2021cgt}. We can go back to our operator basis by using field redefinitions. This is achieved either by extending the operator relations in ref.~\cite{Helset:2022pde} to include fermionic corrections, or by using automated tools such as \texttt{Matchete} \cite{Fuentes-Martin:2022jrf}. We used both methods as a cross-check of our computation, finding agreement in the final result. The RGE results are given in \cref{sec:RGEresults}.

%%%%%%%%%%%%%%%%%%%%%%%%%%%%%%%%%%%%%%%%%%%%%%%%%%%%%%%
\section{Conclusion}\label{sec:Conclusion}
%%%%%%%%%%%%%%%%%%%%%%%%%%%%%%%%%%%%%%%%%%%%%%%%%%%%%%%

Using the geometry of field space, we have extended the one-loop renormalization formula for a general EFT to include mixed boson-fermion graphs. By keeping covariance at every stage of the computation, we have shown that the result depends on geometric quantities such as the field-space curvature. 
As an application of this result, we computed the RGEs of specific two-fermion dimension-eight operators arising from other dimension-eight operators in the SMEFT. 

There are several directions in which this result can be generalized. One extension is to push the computations to two-loop order. While this has already been achieved for a scalar effective field theory \cite{Jenkins:2023bls}, a fully geometric two-loop formula for the RGEs of EFTs involving scalars, fermions, and gauge bosons is still lacking. Additionally, several classes of operators still resist a geometric formulation. In particular, higher-derivative operators pose a challenge to the standard geometric picture, since field redefinitions involving derivatives mix operators with different number of derivatives without affecting physical observables. At tree-level, various modifications to the framework have been attempted to address this issue 
\cite{Cheung:2022vnd,Alminawi:2023qtf,Craig:2023hhp,Craig:2023wni,Cohen:2024bml}. The situation is even more acute at loop level, because any unification of derivative and nonderivative operators will be spoiled by the loop integration. 

While this work represents a significant advance, achieving a fully unified geometric formulation remains an open challenge for future works.

\subsection*{Acknowledgments}

We thank Aneesh Manohar for helpful discussions.
JP is supported by
the U.S. Department of Energy (DOE) under award numbers DE-SC0009919. 
CHS is supported by the Yushan Young Scholarship award number 112V1039 from the Ministry of Education (MOE) of Taiwan, and also by the National Science and Technology Council (NSTC) grant number 114L7329.
BA is supported by the DOE grant number DE--SC0011784 and NSF grant number OAC--2103889 as well as the Fermi National Accelerator Laboratory (Fermilab). This manuscript has been authored by Fermi Forward Discovery Group, LLC under Contract No. 89243024CSC000002 with the U.S. Department of Energy, Office of Science, Office of High Energy Physics. 
This work was performed in part at the Aspen Center for Physics, with support for BA\ by a grant from the Simons Foundation (1161654, Troyer).

%%%%%%%%%%%%%%%%%%%%%%%%%%%%%%%%%%%%%%%%%%%%%%%%%%
\appendix
%%%%%%%%%%%%%%%%%%%%%%%%%%%%%%%%%%%%%%%%%%%%%%%%%%

%%%%%%%%%%%%%%%%%%%%%%%%%%%%%%%%%%%%%%%%%%%%%%%%%%%%%%%
\section{Fermion-scalar supergeometry}\label{sec:Geometry}
%%%%%%%%%%%%%%%%%%%%%%%%%%%%%%%%%%%%%%%%%%%%%%%%%%%%%%%

In this appendix, we will give more details about the supergeometry. Let us pick up the discussion of the scalar-fermion theory in \cref{eq:Lagrsc}. From this Lagrangian, we can extract the metric in \cref{eq:supermetric}, which we reproduce here,
\begin{align} \label{eq:SFmetricfull}
    _a\bar g_{b}(\Phi) 
     = \begin{pmatrix}
        h_{IJ} & \;\;\;(\bar{\psi} \omega^-)_{rI}\;\;\; & (\omega^+ \psi )_{\bar{r}I} \\
        -(\bar{\psi} \omega^-)_{pJ}  & 0 & k_{\bar r p} \\
       -(\omega^+ \psi )_{\bar{p}J} & - k_{\bar p r} & 0
    \end{pmatrix}\,,
\end{align}
where $\omega^{\pm}$ is defined in eq.~\eqref{eq:omegapm}.
We can then invert the metric using $_a\bar{g}_b\bar g^{bc}=\,_a\delta^c$ which gives
\begin{align}
    ^a\bar g^{b}=\bar g^{ab}
     = \begin{pmatrix}
        h^{IJ} & \;\;\;-(\omega^+ \psi )^{rI}\;\;\; & (\bar{\psi} \omega^-)^{\bar r I} \\
        -(\omega^+ \psi )^{pJ} & \Omega_{++}^{pr} & -k^{p \bar r }-\Omega_{+-}^{p\bar r} \\
       (\bar{\psi} \omega^-)^{\bar{p}J} &  k^{r \bar p }+\Omega_{+-}^{r \bar p} & \Omega_{--}^{\bar p \bar r}
    \end{pmatrix}\,,
\end{align}
where 
\begin{align}
    (\omega^{+}\psi)^{pJ} &=  (\omega^{+}\psi)_{I\bar r} k^{p\bar r} h^{IJ} \,,\\
    (\bar\psi\omega^{-})^{\bar pJ} &=  (\bar\psi\omega^{-})_{I r} k^{r\bar p} h^{IJ} \,,\\
    \Omega^{pr}_{++} &= h_{IJ} (\omega^+ \psi)^{p I} (\omega^+ \psi)^{r J}  \,,\\
    \Omega^{p \bar r}_{+-} &= h_{IJ} (\omega^+ \psi)^{p I} (\bar \psi \omega^- )^{\bar r J}  \,,\\
    \Omega^{\bar p \bar r}_{--} &= h_{IJ} (\bar \psi \omega^- )^{\bar p I} (\bar \psi \omega^- )^{\bar r J}  \,.
\end{align}
One property that we have used repeatedly when inverting the metric is that the functions $k_{\bar p r}$ and $\omega_{\bar p r I}$ are diagonal in chirality space. From this, we extract the identity that products such as 
\begin{equation}\label{eq:chiralVanishIdentity}
    k^{p \bar r}  (\bar \psi \omega^-)_{p I} \Gamma_{\mu_1 \dots \mu_{2n}} (\omega^+ \psi)_{\bar r J} = 0
\end{equation}
vanish for any even number of gamma matrices, $\Gamma_{\mu_1 \dots \mu_{2n}}$, including  the scalar bilinear $n=0$.

Gauge bosons can be added to the mix. We introduce the gauge fields $A^{A}_{\mu}$ as additional coordinates in the supermanifold~\cite{Helset:2022tlf,Helset:2022pde}. The full set of supermanifold coordinates is
\begin{align}
    \Phi^{a} = 
    \begin{pmatrix}
        \phi^{I} \\
        A^{A}_{\mu} \\
        \psi^{p} \\
        \bar \psi^{\bar p}
    \end{pmatrix} \,.
\end{align}
The supermetric is then augmented to include the gauge-field components. The combined scalar-gauge-fermion supermetric is given by
\begin{align}
    {}_a\bar g_{b}(\Phi) 
     = \begin{pmatrix}
        h_{IJ} & 0 & (\bar{\psi} \omega^- )_{rI} & (\omega^+ \psi )_{\bar{r}I} \\
        0 & -g_{AB}\eta_{\mu_A\mu_B} & 0 & 0 \\
        -(\bar{\psi} \omega^- )_{pJ} & 0 & 0 & k_{\bar r p} \\
       -(\omega^+ \psi )_{\bar{p}J} & 0 & - k_{\bar p r} & 0
    \end{pmatrix}\,,
\end{align}
where $g_{AB}$ is the gauge-field metric. To obtain this form of the metric, we have used a geometric gauge-fixing term~\cite{Helset:2018fgq}. The gauge fields are bosonic, so their indices carry Grassmann parity zero. 
This supermetric now encapsulates all the fields in the theory---scalars, fermions, and gauge bosons---within a unified framework. The inverse supermetric is
\begin{align}
    \bar g^{ab}
     = \begin{pmatrix}
        h^{IJ} & 0 & -(\omega^+ \psi )^{rI} & (\bar{\psi} \omega^- )^{\bar r I} \\
        0 & -g^{AB}\eta^{\mu_A\mu_B} & 0 & 0 \\
        -(\omega^+ \psi )^{pJ} & 0 & \Omega_{++}^{pr} & -k^{p \bar r } - \Omega_{+-}^{p\bar r} \\
       (\bar{\psi} \omega^- )^{\bar{p}J} & 0 &  k^{r \bar p } + \Omega_{+-}^{r \bar p} & \Omega_{--}^{\bar p \bar r}
    \end{pmatrix}\,,
\end{align}
where $g^{AB}$ is the inverse of the gauge metric $g_{AB}$, and the other terms are as previously defined.

From the supermetric, we can derive various descendant geometric quantities. We start with the Christoffel symbol, which is defined as 
\begin{equation}
\bar\Gamma_{abc} = \frac{1}{2} \left[ \bar g_{ab,c} + (-1)^{\epsilon_b\epsilon_c}\bar g_{ac,b} - (-1)^{\epsilon_a(\epsilon_b+\epsilon_c)}\bar g_{bc,a} \right], \quad\bar{\Gamma}^a_{bc} = (-1)^{\epsilon_d}g^{ad}\Gamma_{dbc}.
\end{equation}
Note that the connection is supersymmetric in the two last indices,
\begin{align}
    \bar \Gamma_{abc} = (-1)^{\epsilon_b\epsilon_c} \bar \Gamma_{acb} .
\end{align}
We can verify that this connection satisfies metric compatibility~\cite{DeWitt:2012mdz},
\begin{equation}
   \bar g_{ab;c}= \bar g_{ab,c}-(-1)^{\epsilon_b(\epsilon_a+\epsilon_d)}\bar g_{db}\bar\Gamma^{d}_{ac}-\bar g_{ad}\bar\Gamma^{d}_{bc}=0 \,.
\end{equation}
Explicitly, the various scalar-fermion components of the Christoffel symbols are
\begin{align}\label{eq:ChristoffelFermion}
\bar \Gamma^{I}_{JK} &= \frac{1}{2} h^{IL} \left( h_{LK,J} + h_{JL,K} - h_{JK,L} \right) 
, \\
%%%%%%%%%%%%%%%%%%%%%%%%
\bar\Gamma^{p}_{JK} &= \frac{1}{2} k^{p\bar q}  (\omega^+ \psi )_{\bar q \lbrace J;K\rbrace} \,
, \\
%%%%%%%%%%%%%%%%%%%%%%%%
\bar\Gamma^{\bar p}_{JK} &= - \frac{1}{2} k^{q \bar p }  (\bar{\psi} \omega^-)_{q \lbrace J;K\rbrace}  \,
, \\
%%%%%%%%%%%%%%%%%%%%%%%%
\bar\Gamma^{p}_{rK} &= -\frac{1}{2} (\omega^+ \psi )^{pI} (\bar{\psi} \omega^- )_{r [I;K]}  +  \left(k^{p\bar s} + \Omega_{+-}^{p\bar s}\right)\omega^+_{\bar s r K} \\
%%%%%%%%%%%%%%%%%%%%%%%%
\bar \Gamma^{\bar p}_{\bar rK} &= \frac{1}{2} (\bar{\psi} \omega^-)^{\bar pI} (\omega^+ \psi)_{\bar r[I;K]}  - \left(k^{\bar p s} + \Omega_{+-}^{s \bar p }\right)\omega^-_{\bar r s K}, \\
%%%%%%%%%%%%%%%%%%%%%%%%
\bar\Gamma^{\bar p}_{r K} &= \frac{1}{2} (\bar{\psi} \omega^-)^{\bar pI} (\bar{\psi} \omega^-)_{ r[I;K]}  -\Omega_{--}^{\bar p  \bar s}\omega^+_{ \bar s r K}, \\
%%%%%%%%%%%%%%%%%%%%%%%%
\bar \Gamma^{ p}_{\bar rK} &= -\frac{1}{2} (\omega^+ \psi)^{ pI} (\omega^+ \psi)_{ \bar r[I;K]}  -\Omega_{++}^{ p s}\omega^-_{ \bar r  s K}, \\
%%%%%%%%%%%%%%%%%%%%%%%%
\bar \Gamma^{I}_{pK} &= \frac{1}{2} h^{IJ} (\bar{\psi} \omega^- )_{p[J;K]} - (\bar{\psi} \omega^-)^{\bar s I}\omega^{+}_{\bar s p K}, \\
%%%%%%%%%%%%%%%%%%%%%%%%
\bar \Gamma^{I}_{\bar pK} &=  \frac{1}{2} h^{IJ} (\omega^+ \psi)_{\bar p[J;K]}  +(\omega^+ \psi)^{ s I}\omega^{-}_{ \bar p s K}, 
%%%%%%%%%%%%%%%%%%%%%%%%
\end{align}
where $F_{[a;b]}=\nabla_bF_a-\nabla_aF_b$, and $F_{\lbrace a;b\rbrace}=\nabla_bF_a+\nabla_aF_b$. Here, $\nabla$ (with no bar) uses only the scalar Christoffel symbol for the scalar submanifold defined by $h_{IJ}$. The covariant derivative $\bar\nabla$ uses the Christoffel symbols for the full supermanifold defined above. In addition, the scalar-gauge Christoffel symbol is given in ref.~\cite{Helset:2022tlf}. There is no nonzero connection with a mixture of fermion and gauge indices due to the block-diagonal structure of the supermetric.

Next, the Riemann curvature is given by
\begin{equation}
    \bar R_{abcd} = g_{ae}R^{e}_{bcd} = \bar g_{ae}\left(-\bar\Gamma^e_{bc,d}+(-1)^{cd}\bar\Gamma^e_{bd,c}+(-1)^{c(b+f)}\bar\Gamma^e_{cf}\bar\Gamma^f_{bd}-(-1)^{d(b+c+f)}\bar\Gamma^e_{fd}\bar\Gamma^f_{bc}\right) \,.
\end{equation}
We can evaluate this Riemann curvature at the vacuum expectation value (VEV) for the fields. For example, the Riemann curvature with two scalar and two fermion indices is \cite{Assi:2023zid}
\begin{align}
\label{eq:CurvatureFermion}
    \bar R_{\bar p r IJ} &= \omega_{\bar p r  J,I} + \omega^+_{\bar ps I} k^{s\bar t}  \omega^-_{\bar tr J}  - \left( I \leftrightarrow J \right) \,.
\end{align}
As expected, the Riemann curvature satisfies all the usual symmetry and Bianchi identities appropriate for a curvature on a supermanifold \cite{Arnowitt:1975bd}.

%%%%%%%%%%%%%%%%%%%%%%%%%%%%%%%%%%%%%%%%%%%%%%%%%%%%%%%
\section{Field-basis covariance}\label{sec:Covariance}
%%%%%%%%%%%%%%%%%%%%%%%%%%%%%%%%%%%%%%%%%%%%%%%%%%%%%%%

Here we will demonstrate that the first and second variations of the action are covariant under changes of field basis. As a result, the full renormalization counterterm Lagrangian will also be covariant.

The class of field redefinitions we will consider are of the form
\begin{align}
    \label{eq:phiTransform}
    \phi^{I} &\rightarrow \phi^{\prime I}(\phi) \,, \\
    \label{eq:psiTransform}
    \psi^{r} &\rightarrow \psi^{\prime r}(\phi,\psi) = R(\phi)^{r}_{\;s}\psi^{s} \,, \\
    \label{eq:barpsiTransform}
    \bar\psi^{\bar p} &\rightarrow \bar \psi^{\prime \bar p}(\phi,\bar\psi) = \bar\psi^{\bar s}(R^{\dagger}(\phi))^{\;\bar r}_{\bar s} \,.
\end{align}
The building blocks of the scalar-fermion metric transforms as
\begin{align}
    h_{IJ} &\rightarrow \left(\frac{\delta \phi^{K}}{\delta \phi^{\prime I}} \right) h_{KL} \left(\frac{\delta \phi^{L}}{\delta \phi^{\prime J}} \right) \,, \\
    k_{\bar pr} &\rightarrow \left[ R^{\dagger-1}k R^{-1} \right]_{\bar pr} \,, \\
    \left[\omega^{\pm}_{I}\right]_{\bar pr} &\rightarrow
    \left[R^{\dagger-1}\omega^{\pm}_{J}R^{-1}
    + \frac{1\pm 1}{2} R^{\dagger-1}kR^{-1}_{,J} - \frac{1\mp 1}{2} R^{\dagger-1}_{,J}kR^{-1}
    \right]_{\bar pr} \left(\frac{\delta \phi^{J}}{\delta \phi^{\prime I}} \right) \,.
\end{align}
We want to identify objects with simple transformation properties, i.e., tensors. To do so, we first need to establish how tensors transform. We group the scalar and fermion coordinates into a multiplet as in \cref{eq:chart_scalarfermion},
\begin{align}
    \Phi^{a} = 
    \begin{pmatrix}
        \phi^{I} \\
        \psi^{p} \\
        \bar \psi^{\bar p}
    \end{pmatrix} \,.
\end{align}
A vector transformation for the combined scalar-fermion field space is given by the matrix
\begin{align}
    \left( \frac{\delta \Phi^{\prime a}}{\delta \Phi^{b}} \right) = 
    \begin{pmatrix}
        \tfrac{\delta \phi^{\prime I}}{\delta \phi^{J}} & 0 & 0 \\
        (R\psi)^{p}_{\;,J} & R^{p}_{\;r} & 0 \\
        (\bar\psi R^\dagger)^{\bar p}_{\;,J} & 0 & (R^{\dagger})_{\bar r}^{\; \bar p} 
    \end{pmatrix} \,,
\end{align} 
and the inverse matrix
\begin{align}
    \left( \frac{\delta \Phi^{a}}{\delta \Phi^{\prime b}} \right) = 
    \begin{pmatrix}
        \tfrac{\delta \phi^{I}}{\delta \phi^{\prime J}} & 0 & 0 \\
        -(R^{-1}R_{,K}\psi)^{p} \tfrac{\delta \phi^{K}}{\delta \phi^{\prime J}} & (R^{-1})^{p}_{\;r} & 0 \\
        -(\bar\psi R^{\dagger}_{,K}R^{\dagger-1})^{\bar p} \tfrac{\delta \phi^{K}}{\delta \phi^{\prime J}} & 0 & (R^{\dagger-1})_{\bar r}^{\; \bar p}
    \end{pmatrix} \,.
\end{align} 
On the scalar field-space geometry, $(D_{\mu}\phi)^{I}$ is a vector. When upgrading the vector to the scalar-fermion manifold, it retains this property,
\begin{align}
    (D_{\mu}\Phi^{a}) = \begin{pmatrix}
        (D_{\mu}\phi^{I}) \\
        (D_{\mu}\psi^{p}) \\
        (D_{\mu}\bar\psi^{\bar p})
    \end{pmatrix} \,,
    \qquad 
    (D_{\mu}\Phi^{a})\rightarrow
    (D_{\mu}\Phi^{\prime a}) =
    \left( \frac{\delta \Phi^{\prime a}}{\delta \Phi^{b}} \right)(D_{\mu}\Phi^{b}) \,.
\end{align}

There is other vectors that we can easily construct, namely the fermion fields themselves. Here we see the fermion fields serving double duty, both as coordinates on the supermanifold and as vectors. Concretely, we can construct the vectors
\begin{align}
    \psi^{a} = 
    \begin{pmatrix}
        0 \\
        \psi^{p} \\
        0
    \end{pmatrix} \,, \qquad 
    \bar \psi^{a} = 
    \begin{pmatrix}
        0 \\
        0 \\
        \bar \psi^{\bar p}
    \end{pmatrix} \,.
\end{align} 
One can verify that their transformation properties in \cref{eq:psiTransform,eq:barpsiTransform} are consistent with them transforming as vectors. From this, we can construct tensors such as $\bar \nabla_{I} \psi^{p}$ and $\bar \nabla_{I} \bar\psi^{\bar p}$, as we did in \cref{eq:nablaPsi,eq:nablaPsiBar}. 

In \cref{eq:defGeomCovD}, we defined derivative operators that preserve the transformation properties of the vectors they act on. Explicitly, the derivative of the fermion fields are still vectors;
\begin{align}
    \label{eq:DphiTransform}
    (\cD_{\mu} \psi^{p}) &\rightarrow R^{p}_{\; r} (\cD_{\mu} \psi^{r}) \,, \\
    \label{eq:DphibarTransform}
    (\cD_{\mu} \bar\psi^{\bar p}) &\rightarrow  (\cD_{\mu} \bar \psi^{\bar r}) (R^{\dagger})^{\; \bar p}_{\bar r} \,.
\end{align}

The fluctuation field also transforms as a contravariant vector
\begin{align}
    \eta^a = \begin{pmatrix}
        \eta^{I} \\
        \chi^{p} \\
        \bar \chi^{\bar p}
    \end{pmatrix}
    \rightarrow
    \begin{pmatrix}
        \tfrac{\delta \phi^{\prime I}}{\delta \phi^{J}} & 0 & 0 \\
        (R\psi)^{p}_{\;,J} & R^{p}_{\;r} & 0 \\
        (\bar\psi R)^{\bar p}_{\;,J} & 0 & (R^{\dagger})_{\bar r}^{\; \bar p}
    \end{pmatrix}
    \begin{pmatrix}
        \eta^{J} \\
        \chi^{r} \\
        \bar \chi^{\bar r}
    \end{pmatrix} \,.
\end{align}
We use $\eta$ with lower-case superindices for the superfield fluctuation and $\eta$ with upper-case indices for scalar fluctuation. 
In contrast, the partial derivative transforms as a covariant vector
\begin{align}
    \frac{\partial}{\partial \Phi^{a}} \rightarrow \left( \frac{\delta \Phi^{b}}{\delta \Phi^{\prime a}} \right) \frac{\partial}{\partial \Phi^{b}} \,.
\end{align}
We can build an invariant combination from the partial derivative and the fluctuations
\begin{align}
    \label{eq:flucDer}
    \eta^{a} \frac{\partial}{\partial \Phi^{a}} \rightarrow \eta^{\prime a} \frac{\partial}{\partial \Phi^{\prime a}} = \eta^{a} \frac{\partial}{\partial \Phi^{a}} . 
\end{align}
This generalizes to $\eta^{a} \bar\nabla_a$ when it acts on tensors.

In the first variations, the fermion mass term is invariant because the operator in \cref{eq:flucDer} is invariant,
\begin{align}
    \delta S \supset&
    \int d^{4}x \bigg\{ - \bar \psi^{\bar p} \cM_{\bar p r} \psi^{r} \bigg\}\;_{,I} \eta^{I} +
    \int d^{4}x \bigg\{ - \bar \psi^{\bar p} \cM_{\bar p r} \psi^{r} \bigg\}\;_{,s} \chi^{s} + 
    \int d^{4}x \bigg\{ - \bar \psi^{\bar p} \cM_{\bar p r} \psi^{r} \bigg\}\;_{,\bar s} \bar\chi^{\bar s}
    \nonumber \\ &
    = \int d^{4}x \bigg\{ - \bar \psi^{\bar p} \cM_{\bar p r} \psi^{r} \bigg\}\;_{,a} \eta^{a}
    \rightarrow 
    \int d^{4}x \bigg\{ - \bar \psi^{\bar p} \cM_{\bar p r} \psi^{r} \bigg\}\;_{,a} \eta^{a} \,.
\end{align}
However, it is more involved to check the invariance of the fermion kinetic term in the first variations. From the fermion fluctuations we get 
\begin{align}
    \label{eq:covarianceKinLchi}
    \delta_{\chi} S &\supset \int d^{4}x \bigg\{
    -i \left(( \cD_{\mu}\bar\psi_r)\gamma^{\mu} \psi^{r} \right) 
    \bigg\}\;_{,s} \chi^{s}
    \nonumber \\ 
    &\rightarrow 
    \int d^{4}x \bigg\{
    -i \left(( \cD_{\mu}\bar\psi_r)\gamma^{\mu} \psi^{r} \right) 
    \bigg\}\;_{,t} (R^{-1})^{t}_{\;s}  \left[ R^{s}_{\;q} \chi^{q} + (R_{,J}\psi)^{s} \eta^{J} \right] \,.
\end{align}
We also need a term from the scalar fluctuation,
\begin{align}
    \label{eq:covarianceKinLeta}
    \delta_{\eta} S &\supset \int d^{4}x \bigg\{
    -i \left(( \cD_{\mu}\bar\psi_r)\gamma^{\mu} \bar \nabla_{I}\psi^{r} \right) 
    \bigg\} \eta^{I} 
    \nonumber \\
    &\rightarrow 
    \int d^{4}x \bigg\{
    -i \left(( \cD_{\mu}\bar\psi_t) (R^{-1})^{t}_{\;r} \gamma^{\mu} \left[\tfrac{\delta\phi^{K}}{\delta\phi^{\prime I}} R^{r}_{\;s}\bar \nabla_{K}\psi^{s} - \tfrac{\delta\phi^{K}}{\delta\phi^{\prime I}} (R_{,K}\psi)^{r}\right]\right) 
    \bigg\} \tfrac{\delta\phi^{\prime I}}{\delta\phi^{J}} \eta^{J} \,.
\end{align}
The combination of \cref{eq:covarianceKinLchi,eq:covarianceKinLeta} is invariant. 

To make the invariance of the first variation manifest, we reorganize it as follows
\begin{align}
      \delta S =& \delta_\zeta S +
     \int d^{4}x \bigg\{ - \bar \psi^{\bar p} (\cM_{\bar p r} -\sigma_{\mu\nu}F^{\mu\nu}_A T_{\bar p r}^A )\psi^{r} \bigg\}\;_{,a} \eta^{a}
     \\
     &  + \int d^{4}x \; i \bigg\{\bar\nabla_a \bar \psi_r \gamma^\mu (\cD_\mu \psi^r )
     - (\cD_\mu \bar \psi^{\bar p}) \gamma^\mu \bar\nabla_a \psi_{\bar p}   
     + R_{ab \bar p r} (D_\mu \Phi)^b (\bar \psi^{\bar p }\gamma^\mu \psi^r) \bigg\} \eta^{a} \,. \nonumber 
\end{align}
The first line is manifestly invariant, while the second line is invariant because of the transformation property of
\begin{align}
    \eta^{a} \bar \nabla_{a}\chi_{\bar p} \rightarrow \eta^{\prime a} \bar \nabla^{\prime}_{a}\chi^{\prime}_{\bar p} = (R^{\dagger-1})_{\bar p}^{\;\; \bar r}\eta^{a} \bar \nabla_{a}\chi_{\bar r} \,,
\end{align}
combined with the transformation in \cref{eq:DphibarTransform}.

Lastly, we turn to the second variation of the action. Many terms are obviously covariant under field redefinitions. Let us focus on the most intricate term, namely the fermion kinetic term. The second variations contain
\begin{align}
    \nabla_{\eta\eta} S_{\rm fermion} 
    \supset& \frac{1}{2}\int d^{4}x \bigg\{
    i \left((\bar \nabla_{I}\bar \nabla_{J}\bar\psi_{r}) \gamma^{\mu} (\cD_{\mu} \psi^{r}) \right)
    - i \left((\cD_{\mu} \bar\psi_{r}) \gamma^{\mu} (\bar \nabla_{I}\bar \nabla_{J}\psi^{r}) \right)
    \nonumber \\ &  \qquad\quad
    + i \left((\bar \nabla_{I}\bar\psi_{r}) \gamma^{\mu} (\cD_{\mu} \bar \nabla_{J}\psi^{r}) \right)
    - i \left((\cD_{\mu} \bar \nabla_{J}\bar\psi_{r}) \gamma^{\mu} (\bar \nabla_{I}\psi^{r}) \right)
    \bigg\}\; \eta^I \eta^{J}
     \\
    \nabla_{\eta\chi} S_{\rm fermion} 
    \supset&
    \int d^4x \bigg\{ - i \left((\cD_{\mu}\bar\nabla_{I}\bar\psi_{r})\gamma^{\mu}\chi^{r} \right)
    \bigg\} \eta^{I} + \textrm{h.c.}
    \\
    \nabla_{\bar\chi \chi} S_{\rm fermion} 
    \supset&
    \int d^{4}x \left\{ 
    \frac{1}{2} k_{\bar pr} i \left( \bar \chi^{\bar p} \overset{\leftrightarrow}{\slashed \cD} \chi^{r} \right) 
    \right\} \,.
\end{align}
We can combine these terms to 
\begin{align}
    \nabla_{\eta\eta} S_{\rm fermion} 
    \supset& \frac{1}{2}\int d^{4}x \bigg\{
    i \left((\bar \nabla_{a}\bar \nabla_{b}\bar\psi_{r}) \gamma^{\mu} (\cD_{\mu} \psi^{r}) \right)
    - i \left((\cD_{\mu} \bar\psi_{r}) \gamma^{\mu} (\bar \nabla_{a}\bar \nabla_{b}\psi^{r}) \right)
    \nonumber \\ &  \qquad\quad
    + i \left((\bar \nabla_{a}\bar\psi_{r}) \gamma^{\mu} (\cD_{\mu} \bar \nabla_{b}\psi^{r}) \right)
    - i \left((\cD_{\mu} \bar \nabla_{b}\bar\psi_{r}) \gamma^{\mu} (\bar \nabla_{a}\psi^{r}) \right)
    \bigg\}\; \eta^{a} \eta^{b} \,.
\end{align}
This term is explicitly built from tensors, which guarantees that the second variation of the action is covariant under field redefinitions.

%%%%%%%%%%%%%%%%%%%%%%%%%%%%%%%%%%%%%%%%%%%%%%%%%%%%%%%
\section{Operators}\label{sec:Operators}
%%%%%%%%%%%%%%%%%%%%%%%%%%%%%%%%%%%%%%%%%%%%%%%%%%%%%%%

The operator basis we use is given in the following tables. The dimension-six Warsaw basis for the SMEFT \cite{Grzadkowski:2010es} comprises the bosonic %and fermionic 
operators in \cref{dim:6} %and \cref{dim:6_2fermion,dim:6_4fermion}, respectively
. The SMEFT operators at dimension eight are similarly split into bosonic operators in \cref{dim:8} and fermion bilinear operators in \cref{dim:8_fermion}. Our convention for these operators aligns with the operator basis in ref.~\cite{Murphy:2020rsh}, with a few exceptions noted in ref.~\cite{Assi:2023zid}.  
%
%
%
%% TABLE dimension-six bosonic operators
%%%%%%%%%%%%%%%%%%%%%%%%%%%%%%%%%%%%%%%%%%%%%%%%%%%%%%%
\begin{table}[H]
\begin{center}
\begin{minipage}[t]{5cm}
\vspace{-0.5cm}
\renewcommand{\arraystretch}{1.5}
\begin{align*}
\begin{array}{c|c|c}
\multicolumn{3}{c}{\bm{X^3}} \\
\hline
Q_G     & \op{6}{G^3}{}           & f^{\mathscr{A}\mathscr{B}\mathscr{C}} G_\mu^{\mathscr{A}\nu} G_\nu^{\mathscr{B}\rho} G_\rho^{\mathscr{C}\mu}  \\
Q_W      & \op{6}{W^3}{}            & \epsilon^{abc} W_\mu^{a\,\nu} W_\nu^{b\,\rho} W_\rho^{c \, \mu} \\ 
\end{array}
\end{align*}
\renewcommand{\arraystretch}{1.5}
\vspace{-0.5cm}
\begin{align*}
\begin{array}{c|c|c}
\multicolumn{3}{c}{\bm{H^4 D^2} } \\
\hline
Q_{H\Box} & \op{6}{H^4\Box}{} & (H^\dag H)\Box (H^\dag H) \\
Q_{H D} &\op{6}{H^4D^2}{}  & \left(D^\mu H^\dag H\right) \left(H^\dag D_\mu H\right)
\end{array}
\end{align*}
\vspace{-0.5cm}
\renewcommand{\arraystretch}{1.5}
\begin{align*}
\begin{array}{c|c|c}
\multicolumn{3}{c}{\bm{H^6} } \\
\hline
Q_H    & \op{6}{H^6}{}   & (H^\dag H)^3
\end{array}
\end{align*}
\end{minipage}
\hspace{1cm}
\begin{minipage}[t]{2.5cm}
\vspace{-0.5cm}
\renewcommand{\arraystretch}{1.5}
\begin{align*}
\begin{array}{c|c|c}
\multicolumn{3}{c}{\bm{X^2 H^2} } \\
\hline
Q_{H G}  & \op{6}{G^2H^2}{(1)}   & (H^\dag H)\, G^{\mathscr{A}}_{\mu\nu} G^{\mathscr{A}\mu\nu} \\
Q_{H \tilde G}  & \op{6}{G^2H^2}{(2)}   & (H^\dag H)\, G^{\mathscr{A}}_{\mu\nu} \tilde G^{\mathscr{A}\mu\nu} \\
Q_{H W}  &\op{6}{W^2H^2}{(1)}     & (H^\dag H) \, W^a_{\mu\nu} W^{a\mu\nu} \\
Q_{H \tilde W}  &\op{6}{W^2H^2}{(2)}     & (H^\dag H) \, W^a_{\mu\nu} \tilde W^{a\mu\nu} \\
Q_{H B} & \op{6}{B^2H^2}{(1)}      &  (H^\dag H)\, B_{\mu\nu} B^{\mu\nu} \\
Q_{H \tilde B} & \op{6}{B^2H^2}{(2)}      &  (H^\dag H)\, B_{\mu\nu} \tilde B^{\mu\nu} \\
Q_{H WB} &  \op{6}{WBH^2}{(1)}    &  (H^\dag \tau^a H)\, W^a_{\mu\nu} B^{\mu\nu} \\
Q_{H \tilde WB} &  \op{6}{WBH^2}{(2)}    &  (H^\dag \tau^a H)\, \tilde W^a_{\mu\nu} B^{\mu\nu} \\
\end{array}
\end{align*}
\end{minipage}
\end{center}
\caption{\label{dim:6} Bosonic dimension-six operators in the SMEFT. The first column is the notation of ref.~\cite{Grzadkowski:2010es}, and the second column is the notation used in this paper.}
\end{table}
%%%%%%%%%%%%%%%%%%%%%%%%%%%%%%%%%%%%%%%%%%%%%%%%%%%%%%%
%
%
%
%% TABLE dimesion-six two-fermion operators
%%%%%%%%%%%%%%%%%%%%%%%%%%%%%%%%%%%%%%%%%%%%%%%%%%%%%%%
\begin{table}[H]
\begin{center}
\begin{minipage}[t]{5cm}
\vspace{-1cm}
\renewcommand{\arraystretch}{1.5}
\begin{align*}
\begin{array}{c|c|c}
\multicolumn{3}{c}{\bm{\psi^2 H^3 + \textbf{h.c.}}} \\
\hline
Q_{eH}    & \op{6}{\ell e H^3}{}           & (H^{\dagger} H)(\bar \ell_{p} e_{r} H)  \\
Q_{uH}    & \op{6}{q u H^3}{}           & (H^{\dagger} H)(\bar q_{p} u_{r}\tilde H)  \\
Q_{dH}    & \op{6}{q d H^3}{}           & (H^{\dagger} H)(\bar q_{p} d_{r} H)  \\
\end{array}
\end{align*}
\vspace{-.5cm}
\renewcommand{\arraystretch}{1.5}
\begin{align*}
\begin{array}{c|c|c}
\multicolumn{3}{c}{\bm{\psi^2 X H + \textbf{h.c.}}} \\
\hline
Q_{eW} & \op{6}{\ell e W H}{} & (\bar \ell_{p} \sigma^{\mu\nu} \tau^{a} e_{r} H) W^{a}_{\mu\nu} \\
Q_{eB} & \op{6}{\ell e B H}{} & (\bar \ell_{p} \sigma^{\mu\nu}  e_{r} H) B_{\mu\nu} \\
Q_{uG} & \op{6}{q u G H}{} & (\bar q_{p} \sigma^{\mu\nu} T^{\mathscr{A}} u_{r} \tilde H) G^{\mathscr{A}}_{\mu\nu} \\
Q_{uW} & \op{6}{q u W H}{} & (\bar q_{p} \sigma^{\mu\nu} \tau^{a} u_{r} \tilde H) W^{a}_{\mu\nu} \\
Q_{uB} & \op{6}{q u B H}{} & (\bar q_{p} \sigma^{\mu\nu} u_{r} \tilde H) B_{\mu\nu} \\
Q_{dG} & \op{6}{q d G H}{} & (\bar q_{p} \sigma^{\mu\nu} T^{\mathscr{A}} d_{r} H) G^{\mathscr{A}}_{\mu\nu} \\
Q_{dW} & \op{6}{q d W H}{} & (\bar q_{p} \sigma^{\mu\nu} \tau^{a} d_{r} H) W^{a}_{\mu\nu} \\
Q_{dB} & \op{6}{q d B H}{} & (\bar q_{p} \sigma^{\mu\nu} d_{r} H) B_{\mu\nu} \\
\end{array}
\end{align*}
\end{minipage}
\hspace{1cm}
\begin{minipage}[t]{2.5cm}
\vspace{-1cm}
\renewcommand{\arraystretch}{1.5}
\renewcommand{\arraystretch}{1.5}
\begin{align*}
\begin{array}{c|c|c}
\multicolumn{3}{c}{\bm{\psi^2 H^2 D} } \\
\hline
Q_{H \ell}^{(1)}  & \op{6}{\ell^2H^2 D}{(1)}   & (\bar \ell_{p} \gamma^{\mu} \ell_{r})\,(H^\dag i \overset{\leftrightarrow}{D}_{\mu} H)  \\
Q_{H \ell}^{(3)}  & \op{6}{\ell^2H^2 D}{(3)}   & (\bar \ell_{p} \gamma^{\mu} \tau^{a} \ell_{r})\,(H^\dag i \overset{\leftrightarrow}{D}_{\mu}^{a} H)  \\
Q_{H e}  & \op{6}{e^2H^2 D}{}   & (\bar e_{p} \gamma^{\mu} e_{r})\,(H^\dag i \overset{\leftrightarrow}{D}_{\mu} H)  \\
Q_{H q}^{(1)}  & \op{6}{q^2H^2 D}{(1)}   & (\bar q_{p} \gamma^{\mu} q_{r})\,(H^\dag i \overset{\leftrightarrow}{D}_{\mu} H)  \\
Q_{H q}^{(3)}  & \op{6}{q^2H^2 D}{(3)}   & (\bar q_{p} \gamma^{\mu}\tau^a q_{r})\,(H^\dag i \overset{\leftrightarrow}{D}_{\mu}^{a} H)  \\
Q_{H u}  & \op{6}{u^2H^2 D}{}   & (\bar u_{p} \gamma^{\mu} u_{r})\,(H^\dag i \overset{\leftrightarrow}{D}_{\mu} H)  \\
Q_{H d}  & \op{6}{d^2H^2 D}{}   & (\bar d_{p} \gamma^{\mu} d_{r})\,(H^\dag i \overset{\leftrightarrow}{D}_{\mu} H)  \\
Q_{H ud} + \textrm{h.c.}  & \op{6}{udH^2 D}{}   & (\bar u_{p} \gamma^{\mu} d_{r})\, (\tilde H^\dag i D_{\mu} H)  \\
\end{array}
\end{align*}
\end{minipage}
\end{center}
\caption{\label{dim:6_2fermion} Two-fermion dimension-six operators in the SMEFT. The first column is the notation of ref.~\cite{Grzadkowski:2010es}, and the second column is the notation used in this paper.}
\end{table}
\begin{table}[H]
\begin{center}
\begin{minipage}[t]{2.15cm}
\vspace{-0.5cm}
\renewcommand{\arraystretch}{1.5}
\begin{align*}
\begin{array}{c|c}
\multicolumn{2}{c}{\bm{H^8} } \\
\hline
\op{8}{H^8}{} &  (H^\dag H)^4 
\end{array}
\end{align*}
\vspace{-0.5cm}
\renewcommand{\arraystretch}{1.5}
\begin{align*}
\begin{array}{c|c}
\multicolumn{2}{c}{\bm{H^6 D^2} } \\
\hline
\op{8}{H^6 D^2}{(1)}  & (H^{\dag} H)^2 (D_{\mu} H^{\dag} D^{\mu} H) \\
\op{8}{H^6 D^2}{(2)}  & (H^{\dag} H) (H^{\dag} \tau^I H) (D_{\mu} H^{\dag} \tau^I D^{\mu} H)
\end{array}
\end{align*}

\vspace{-0.5cm}
\renewcommand{\arraystretch}{1.5}
\begin{align*}
\begin{array}{c|c}
\multicolumn{2}{c}{\bm{H^4 D^4} } \\
\hline
\op{8}{H^4 D^4}{(1)}  &  (D_{\mu} H^{\dag} D_{\nu} H) (D^{\nu} H^{\dag} D^{\mu} H) \\ 
\op{8}{H^4 D^4}{(2)}  &  (D_{\mu} H^{\dag} D_{\nu} H) (D^{\mu} H^{\dag} D^{\nu} H) \\ 
\op{8}{H^4 D^4}{(3)}  &  (D^{\mu} H^{\dag} D_{\mu} H) (D^{\nu} H^{\dag} D_{\nu} H)
\end{array}
\end{align*}
\vspace{-0.5cm}
\renewcommand{\arraystretch}{1.5}
\begin{align*}
\begin{array}{c|c}
\multicolumn{2}{c}{\bm{X^3 H^2} } \\
\hline
\op{8}{G^3H^2}{(1)}  &  f^{\mathscr{A}\mathscr{B}\mathscr{C}} (H^\dag H) G_{\mu}^{\mathscr{A}\nu} G_{\nu}^{\mathscr{B}\rho} G_{\rho}^{\mathscr{C}\mu} \\
\op{8}{W^3H^2}{(1)}  &  \epsilon^{abc} (H^\dag H) W_{\mu}^{a\,\nu} W_{\nu}^{b\,\rho} W_{\rho}^{c\,\mu} \\
\op{8}{W^2BH^2}{(1)}  &  \epsilon^{abc} (H^\dag \tau^a H) B_{\mu}^{\;\;\,\nu} W_{\nu}^{b\,\rho} W_{\rho}^{c\,\mu} \\
\end{array}
\end{align*}
%%%
\vspace{-0.5cm}
\renewcommand{\arraystretch}{1.5}
\begin{align*}
\begin{array}{c|c}
\multicolumn{2}{c}{\bm{X^2 H^4} } \\
\hline
\op{8}{G^2H^4}{(1)}  & (H^\dag H)^2 G^{\mathscr{A}}_{\mu\nu} G^{\mathscr{A}\mu\nu} \\
\op{8}{G^2H^4}{(2)}  & (H^\dag H)^2 G^{\mathscr{A}}_{\mu\nu} \tilde G^{\mathscr{A}\mu\nu} \\
\op{8}{W^2H^4}{(1)}  & (H^\dag H)^2 W^a_{\mu\nu} W^{a\,\mu\nu} \\
\op{8}{W^2H^4}{(2)}  & (H^\dag H)^2 W^a_{\mu\nu} \tilde W^{a\,\mu\nu} \\
\op{8}{W^2H^4}{(3)}  & (H^\dag \tau^a H) (H^\dag \tau^b H) W^a_{\mu\nu} W^{b\,\mu\nu} \\
\op{8}{W^2H^4}{(4)}  & (H^\dag \tau^a H) (H^\dag \tau^b H) W^a_{\mu\nu} \tilde W^{b\,\mu\nu} \\
\op{8}{WBH^4}{(1)}  &  (H^\dag H) (H^\dag \tau^a H) W^a_{\mu\nu} B^{\mu\nu} \\
\op{8}{WBH^4}{(2)}  &  (H^\dag H) (H^\dag \tau^a H) \tilde W^a_{\mu\nu} B^{\mu\nu} \\
\op{8}{B^2H^4}{(1)}  &  (H^\dag H)^2 B_{\mu\nu} B^{\mu\nu} \\
\op{8}{B^2H^4}{(2)}  &  (H^\dag H)^2 B_{\mu\nu} \tilde B^{\mu\nu} \\
\end{array}
\end{align*}
%%%%
\end{minipage}
\hspace{0.5cm}
\begin{minipage}[t]{4cm}
\iffalse
\vspace{-0.5cm}
\renewcommand{\arraystretch}{1.5}
\begin{align*}
\begin{array}{c|c}
\multicolumn{2}{c}{\bm{X^2 H^4} } \\
\hline
%
\op{8}{G^2H^4}{(1)}  & (H^\dag H)^2 G^{\mathscr{A}}_{\mu\nu} G^{\mathscr{A}\mu\nu} \\
%
\op{8}{G^2H^4}{(2)}  & (H^\dag H)^2 G^{\mathscr{A}}_{\mu\nu} \tilde G^{\mathscr{A}\mu\nu} \\
%
\op{8}{W^2H^4}{(1)}  & (H^\dag H)^2 W^a_{\mu\nu} W^{a\,\mu\nu} \\
%
\op{8}{W^2H^4}{(2)}  & (H^\dag H)^2 W^a_{\mu\nu} \tilde W^{a\,\mu\nu} \\
%
\op{8}{W^2H^4}{(3)}  & (H^\dag \tau^a H) (H^\dag \tau^b H) W^a_{\mu\nu} W^{b\,\mu\nu} \\
%
\op{8}{W^2H^4}{(4)}  & (H^\dag \tau^a H) (H^\dag \tau^b H) W^a_{\mu\nu} \tilde W^{b\,\mu\nu} \\
%
\op{8}{WBH^4}{(1)}  &  (H^\dag H) (H^\dag \tau^a H) W^a_{\mu\nu} B^{\mu\nu} \\
%
\op{8}{WBH^4}{(2)}  &  (H^\dag H) (H^\dag \tau^a H) \tilde W^a_{\mu\nu} B^{\mu\nu} \\
%
\op{8}{B^2H^4}{(1)}  &  (H^\dag H)^2 B_{\mu\nu} B^{\mu\nu} \\
%
\op{8}{B^2H^4}{(2)}  &  (H^\dag H)^2 B_{\mu\nu} \tilde B^{\mu\nu} \\
%
\end{array}
\end{align*}
\fi
\vspace{-0.5cm}
\renewcommand{\arraystretch}{1.5}
\begin{align*}
\begin{array}{c|c}
\multicolumn{2}{c}{\bm{X  H^4 D^2} } \\
\hline
\op{8}{WH^4D^2}{(1)}  & i (H^{\dag} H) (D^{\mu} H^{\dag} \tau^a D^{\nu} H) W_{\mu\nu}^a \\
\op{8}{WH^4D^2}{(2)}  & i (H^{\dag} H) (D^{\mu} H^{\dag} \tau^a D^{\nu} H) \tilde W_{\mu\nu}^a \\
\op{8}{WH^4D^2}{(3)}  & i \epsilon^{abc} (H^{\dag} \tau^a H) (D^{\mu} H^{\dag} \tau^b D^{\nu} H) W_{\mu\nu}^c \\
\op{8}{WH^4D^2}{(4)}  & i \epsilon^{abc} (H^{\dag} \tau^a H) (D^{\mu} H^{\dag} \tau^b D^{\nu} H) \tilde W_{\mu\nu}^c \\
\op{8}{BH^4D^2}{(1)}  & i (H^{\dag} H) (D^{\mu} H^{\dag} D^{\nu} H) B_{\mu\nu} \\
\op{8}{BH^4D^2}{(2)}  & i (H^{\dag} H) (D^{\mu} H^{\dag} D^{\nu} H) \tilde B_{\mu\nu} \\
\end{array}
\end{align*}

\vspace{-0.5cm}
\renewcommand{\arraystretch}{1.5}
\begin{align*}
\begin{array}{c|c}
\multicolumn{2}{c}{\bm{X^2 H^2 D^2} } \\
\hline
\op{8}{G^2H^2D^2}{(1)}  &  (D^{\mu} H^{\dag} D^{\nu} H) G_{\mu\rho}^{\mathscr{A}} G_{\nu}^{\mathscr{A} \rho} \\
\op{8}{G^2H^2D^2}{(2)}  &  (D^{\mu} H^{\dag} D_{\mu} H) G_{\nu\rho}^{\mathscr{A}} G^{\mathscr{A} \nu\rho} \\
\op{8}{W^2H^2D^2}{(1)}  &  (D^{\mu} H^{\dag} D^{\nu} H) W_{\mu\rho}^a W_{\nu}^{a\, \rho} \\
\op{8}{W^2H^2D^2}{(2)}  &  (D^{\mu} H^{\dag} D_{\mu} H) W_{\nu\rho}^a W^{a\, \nu\rho} \\
\op{8}{W^2H^2D^2}{(4)}  &  i \epsilon^{abc} (D^{\mu} H^{\dag} \tau^a D^{\nu} H) W_{\mu\rho}^b W_{\nu}^{c\, \rho} \\
\op{8}{W^2H^2D^2}{(5)}  &  \epsilon^{abc} (D^{\mu} H^{\dag} \tau^a D^{\nu} H) \\  & \times (W_{\mu\rho}^b \tilde W_{\nu}^{c\, \rho} - \tilde W_{\mu\rho}^b W_{\nu}^{c\, \rho}) \\
\op{8}{WBH^2D^2}{(1)}  &  (D^{\mu} H^{\dag} \tau^a D_{\mu} H) B_{\nu\rho} W^{a\, \nu\rho} \\
\op{8}{WBH^2D^2}{(2)}  &  (D^{\mu} H^{\dag} \tau^a D_{\mu} H) B_{\nu\rho} \tilde W^{a\, \nu\rho} \\
\op{8}{WBH^2D^2}{(3)}  &  i (D^{\mu} H^{\dag} \tau^a D^{\nu} H) \\
 &  \times (B_{\mu\rho} W_{\nu}^{a\, \rho} - B_{\nu\rho} W_{\mu}^{a\,\rho}) \\
\op{8}{WBH^2D^2}{(4)}  &  (D^{\mu} H^{\dag} \tau^a D^{\nu} H) \\
  & \times (B_{\mu\rho} W_{\nu}^{a\, \rho} + B_{\nu\rho} W_{\mu}^{a\,\rho}) \\
\op{8}{WBH^2D^2}{(6)}  &  (D^{\mu} H^{\dag} \tau^a D^{\nu} H) \\
  & \times (B_{\mu\rho} \tilde W_{\nu}^{a\, \rho} + B_{\nu\rho} \tilde W_{\mu}^{a\,\rho}) \\
\op{8}{B^2H^2D^2}{(1)}  &  (D^{\mu} H^{\dag} D^{\nu} H) B_{\mu\rho} B_{\nu}^{\,\,\,\rho} \\
\op{8}{B^2H^2D^2}{(2)}  &  (D^{\mu} H^{\dag} D_{\mu} H) B_{\nu\rho} B^{\nu\rho} \\
\end{array}
\end{align*}

\end{minipage}
\hspace{-1.5cm}
%
%\end{tabular}
\end{center}
\caption{\label{dim:8} Bosonic dimension-eight operators in the SMEFT.  The $XH^4D^2$ operators have a factor of $i$ relative to ref.~\cite{Murphy:2020rsh} to make them hermitian. There are also $X^4$ operators which have not been listed.}
\end{table}
%%%%%%%%%%%%%%%%%%%%%%%%%%%%%%%%%%%%%%%%%%%%%%%%%%%%%%%

%% TABLE two-fermion dim 8
%%
%%%%%%%%%%%%%%%%%%%%%%%%%%%%%%%%%%%%%%%%%%%%%%%%%%%%%%%
\begin{table}[H]
\begin{center}
\begin{minipage}[t]{4.5cm}
\vspace{-0.5cm}
\renewcommand{\arraystretch}{1.5}
\begin{align*}
\begin{array}{c|c}
\multicolumn{2}{c}{\bm{\psi^2 H^5 + \textbf{h.c.}}} \\
\hline
  \op{8}{\ell e H^5}{}           & (H^{\dagger} H)^{2} (\bar \ell_{p} e_{r} H)  \\
  \op{8}{q u H^5}{}           & (H^{\dagger} H)^{2} (\bar q_{p} u_{r}\tilde H)  \\
  \op{8}{q d H^5}{}           & (H^{\dagger} H)^{2} (\bar q_{p} d_{r} H)  \\
\end{array}
\end{align*}
\renewcommand{\arraystretch}{1.5}
\begin{align*}
\begin{array}{c|c}
\multicolumn{2}{c}{\bm{\psi^2 X H^{3} + \textbf{h.c.}}} \\
\hline
 \op{8}{\ell e W H^3}{(1)} & (\bar \ell_{p} \sigma^{\mu\nu} \tau^{a} e_{r} H) (H^\dag H) W^{a}_{\mu\nu} \\
 \op{8}{\ell e W H^3}{(2)} & (\bar \ell_{p} \sigma^{\mu\nu} e_{r} H) (H^\dag \tau^a H) W^{a}_{\mu\nu} \\
 \op{8}{\ell e B H^3}{} & (\bar \ell_{p} \sigma^{\mu\nu}  e_{r} H) (H^\dag H)B_{\mu\nu} \\
 \op{8}{q u G H^3}{} & (\bar q_{p} \sigma^{\mu\nu} T^{\mathscr{A}} u_{r} \tilde H) (H^\dag H) G^{\mathscr{A}}_{\mu\nu} \\
 \op{8}{q u W H^3}{(1)} & (\bar q_{p} \sigma^{\mu\nu} \tau^{a} u_{r} \tilde H) (H^\dag H) W^{a}_{\mu\nu} \\
 \op{8}{q u W H^3}{(2)} & (\bar q_{p} \sigma^{\mu\nu} u_{r} \tilde H) (H^\dag \tau^{a} H) W^{a}_{\mu\nu} \\
 \op{8}{q u B H^3}{} & (\bar q_{p} \sigma^{\mu\nu} u_{r} \tilde H) (H^\dag H) B_{\mu\nu} \\
 \op{8}{q d G H^3}{} & (\bar q_{p} \sigma^{\mu\nu} T^{\mathscr{A}} d_{r} H) (H^\dag H) G^{\mathscr{A}}_{\mu\nu} \\
 \op{8}{q d W H^3}{(1)} & (\bar q_{p} \sigma^{\mu\nu} \tau^{a} d_{r} H) (H^\dag H) W^{a}_{\mu\nu} \\
 \op{8}{q d W H^3}{(2)} & (\bar q_{p} \sigma^{\mu\nu} d_{r} H) (H^\dag \tau^{a} H) W^{a}_{\mu\nu} \\
 \op{8}{q d B H^3}{} & (\bar q_{p} \sigma^{\mu\nu} d_{r} H) (H^\dag H) B_{\mu\nu} \\
\end{array}
\end{align*}
\end{minipage}
\hspace{0cm}
\begin{minipage}[t]{3cm}
\vspace{-0.5cm}
\renewcommand{\arraystretch}{1.5}
\renewcommand{\arraystretch}{1.5}
\begin{align*}
\begin{array}{c|c}
\multicolumn{2}{c}{\bm{\psi^2 H^4 D} } \\
\hline
 \op{8}{\ell^2H^4 D}{(1)}   & (\bar \ell_{p} \gamma^{\mu} \ell_{r})\,(H^\dag i \overset{\leftrightarrow}{D}_{\mu} H) (H^\dag H) \\
 \op{8}{\ell^2H^4 D}{(2)}   & (\bar \ell_{p} \gamma^{\mu} \tau^{a} \ell_{r})\,(H^\dag i \overset{\leftrightarrow}{D}_{\mu} H) (H^\dag \tau^{a} H)  \\
 \op{8}{\ell^2H^4 D}{(3)}   & (\bar \ell_{p} \gamma^{\mu} \tau^{a} \ell_{r})\,(H^\dag i \overset{\leftrightarrow}{D}_{\mu}^{a} H) (H^\dag H)  \\
 \op{8}{\ell^2H^4 D}{(4)}   & \epsilon_{abc}(\bar \ell_{p} \gamma^{\mu} \tau^{a} \ell_{r})\,(H^\dag i \overset{\leftrightarrow}{D}_{\mu}^{b} H) (H^\dag \tau^{c} H)  \\
 \op{8}{e^2H^4 D}{}   & (\bar e_{p} \gamma^{\mu} e_{r})\,(H^\dag i \overset{\leftrightarrow}{D}_{\mu} H)  (H^\dag H) \\
 \op{8}{q^2H^4 D}{(1)}   & (\bar q_{p} \gamma^{\mu} q_{r})\,(H^\dag i \overset{\leftrightarrow}{D}_{\mu} H)  (H^\dag H) \\
 \op{8}{q^2H^4 D}{(2)}   & (\bar q_{p} \gamma^{\mu} \tau^{a} q_{r})\,(H^\dag i \overset{\leftrightarrow}{D}_{\mu} H) (H^\dag \tau^{a} H)  \\
 \op{8}{q^2H^4 D}{(3)}   & (\bar q_{p} \gamma^{\mu} \tau^{a} q_{r})\,(H^\dag i \overset{\leftrightarrow}{D}_{\mu}^{a} H) (H^\dag H)  \\
 \op{8}{q^2H^4 D}{(4)}   & \epsilon_{abc}(\bar q_{p} \gamma^{\mu} \tau^{a} q_{r})\,(H^\dag i \overset{\leftrightarrow}{D}_{\mu}^{b} H) (H^\dag \tau^{c} H)  \\
 \op{8}{u^2H^4 D}{}   & (\bar u_{p} \gamma^{\mu} u_{r})\,(H^\dag i \overset{\leftrightarrow}{D}_{\mu} H) (H^\dag H)  \\
 \op{8}{d^2H^4 D}{}   & (\bar d_{p} \gamma^{\mu} d_{r})\,(H^\dag i \overset{\leftrightarrow}{D}_{\mu} H) (H^\dag H)  \\
 \op{8}{udH^4 D}{} + \textrm{h.c.}  & (\bar u_{p} \gamma^{\mu} d_{r})\, (\tilde H^\dag i D_{\mu} H) (H^\dag H)   \\
\end{array}
\end{align*}
\end{minipage}
\end{center}
\caption{\label{dim:8_fermion} Fermionic dimension-eight operators in the SMEFT (not including four-fermion operators or other operators not appearing in the initial Lagrangian). This notation coincides for the most part with the notation used in ref.~\cite{Murphy:2020rsh}.}
\end{table}
%%%%%%%%%%%%%%%%%%%%%%%%%%%%%%%%%%%%%%%%%%%%%%%%%%%%%%%

%%%%%%%%%%%%%%%%%%%%%%%%%%%%%%%%%%%%%%%%%%%%%%%%%%%%%%%
\section{Renormalization Group Evolution in the SMEFT to Dimension Eight}\label{sec:RGEresults}
%%%%%%%%%%%%%%%%%%%%%%%%%%%%%%%%%%%%%%%%%%%%%%%%

We now list the renormalization group equations for two-fermion operators renormalized by boson, fermion, and mixed boson-fermion loops in the SMEFT up to dimension eight. This comes in addition to the RGE results for the bosonic sector~\cite{Chala:2021pll,DasBakshi:2022mwk,AccettulliHuber:2021uoa,Helset:2022pde,Assi:2023zid}. We use the notation $\dot C = 16\pi^2 \mu \tfrac{\rm{d}}{\rm{d}\mu}C$.

%%%%%%%%%%%%%%%%%%%%%%%%%%%%%%%%%%%%%%%%%%%%%%%%
\subsection{Definitions}
%%%%%%%%%%%%%%%%%%%%%%%%%%%%%%%%%%%%%%%%%%%%%%%%

Here we list some combinations of couplings that enter in the RGE results below. 

\subsubsection*{Wavefunction factors}

For the wavefunction factors, we define
\begin{align}
	\gamma_{H}^{(Y)} =& 
	{\rm Tr}\left[ Y^{\dagger}_{e} Y_{e} +  N_{c} Y^{\dagger}_{u} Y_{u} + N_{c} Y^{\dagger}_{d} Y_{d} \right] \,, \\
    \gamma_{e,\bar pr}^{(Y)} =&  
	\left[ Y_{e} Y^{\dagger}_{e} \right]_{\bar pr} \,, \\
    \gamma_{\ell,\bar pr}^{(Y)} =&  
	\frac{1}{2}\left[ Y^{\dagger}_{e} Y_{e} \right]_{\bar pr} \,, \\
    \gamma_{u,\bar pr}^{(Y)} =&  
	\left[ Y_{u} Y^{\dagger}_{u} \right]_{\bar pr} \,, \\
    \gamma_{d,\bar pr}^{(Y)} =&  
	\left[ Y_{d} Y^{\dagger}_{d} \right]_{\bar pr} \,, \\
    \gamma_{q,\bar pr}^{(Y)} =&  
	\frac{1}{2}\left[ Y^{\dagger}_{u} Y_{u} + Y^{\dagger}_{d} Y_{d} \right]_{\bar pr} \,.
\end{align}
These wavefunction factors come in addition to the ones from the gauge interactions.

\subsubsection*{Boson loop}

The boson loops contribute to the renormalization group equations of the fermionic operators in two ways, either directly (through the terms in the bosonic second variations of the action in \cref{eq:variationEtaEta,eq:variationEtaZeta,eq:variationZetaZeta} which contain fermions), or via field redefinitions. For the latter case, we define 
\begin{align}
    \dimEightBosonLoopOne &= \frac{1}{3}   g_1^2 \left(   \coef{8}{H^6D^2}{(1)} + 2 \coef{8}{H^6D^2}{(2)} \right)-  8 g_1^2  \coef{8}{B^2H^4}{(1)}  - 8 g_1 g_2 \coef{8}{WBH^4}{(1)}  \,, \\
    \dimEightBosonLoopTwo &= \frac{1}{6}g_2^2\coef{8}{H^6D^2}{(2)} -  g_1 g_2 \coef{8}{WBH^4}{(1)}  - 10 g_2^2   \coef{8}{W^2H^4}{(3)}   \,, \\
    \dimEightBosonLoopThree &= \frac{1}{6} g_2^2 \left( \coef{8}{H^6D^2}{(1)} - \coef{8}{H^6D^2}{(2)} \right) -  g_1 g_2 \coef{8}{WBH^4}{(1)}  -2 g_2^2 \left( 2 \coef{8}{W^2H^4}{(1)} - \coef{8}{W^2H^4}{(3)}  \right)  \,, \\
    \dimEightBosonLoopYuk &= \frac{1}{12}\left( - 24 \lambda \left( \coef{8}{H^6D^2}{(1)} + 4 \coef{8}{H^6D^2}{(2)} \right) + 2 \coef{8}{H^6D^2}{(2)} \left( 2g_1^2 - 27g_2^2 \right) + 5 \coef{8}{H^6D^2}{(1)} \left( g_1^2 + 7g_2^2 \right) \right) \nonumber \\
    &\quad - 4\left( g_1^2\coef{8}{B^2H^4}{(1)} + g_1 g_2\coef{8}{WBH^4}{(1)} + g_2^2 \left( 3\coef{8}{W^2H^4}{(1)}  + \coef{8}{W^2H^4}{(3)}    \right)\right) \,.
\end{align}
The operator relations in ref.~\cite{Helset:2022pde}, which are valid in the bosonic sector, must be amended to include the missing fermion operators.

\subsubsection*{Fermion loop}

For the fermionic loops, the contributions to the fermion RGEs come either from four-fermion operators or appear after a field redefinition. It is convenient to group the couplings in the same way as in ref.~\cite{Assi:2023zid}. We use the same notation (with the same numerical labels), but we only list the coefficient that enter the fermion RGEs.
\begin{align}
%	\kappaHypKinSix =& \left[ y_{e} \; \coef{6}{\underset{tt}{e^2 H^2 D}}{} + 2 y_{\ell} \; \coef{6}{\underset{tt}{\ell^2 H^2 D}}{(1)}  + N_{c} y_{u}  \; \coef{6}{\underset{tt}{u^2 H^2 D}}{} + N_{c} y_{d}  \; \coef{6}{\underset{tt}{d^2 H^2 D}}{}  + 2 N_{c} y_{q}  \;  \coef{6}{\underset{tt}{q^2 H^2 D}}{(1)} \right] \,,  \\
    \kappaHypKinEight =& \left[ y_{e} \; \coef{8}{\underset{tt}{e^2 H^4 D}}{} + 2 y_{\ell} \; \coef{8}{\underset{tt}{\ell^2 H^4 D}}{(1)} + N_{c} y_{u}  \; \coef{8}{\underset{tt}{u^2 H^4 D}}{} + N_{c} y_{d}  \; \coef{8}{\underset{tt}{d^2 H^4 D}}{}  + 2 N_{c} y_{q}  \; \coef{8}{\underset{tt}{q^2 H^4 D}}{(1)}  \right] \,,  \\
%    \kappaPauliKinSix =& \left[ \coef{6}{\underset{tt}{\ell^2 H^2 D}}{(3)}  + N_{c} \; \coef{6}{\underset{tt}{q^2 H^2 D}}{(3)} \right] \,, \\
	\kappaPauliKinEight =& \left[ \coef{8}{\underset{tt}{\ell^2 H^4 D}}{(3)}  + N_{c} \; \coef{8}{\underset{tt}{q^2 H^4 D}}{(3)}  \right] \,, \\
    \kappaPauliKinEightTwo =& \left[ \coef{8}{\underset{tt}{\ell^2 H^4 D}}{(2)}  + N_{c} \; \coef{8}{\underset{tt}{q^2 H^4 D}}{(2)}  \right] \,, \\
    \kappaPauliKinEightFour =& \left[ \coef{8}{\underset{tt}{\ell^2 H^4 D}}{(4)}  + N_{c} \; \coef{8}{\underset{tt}{q^2 H^4 D}}{(4)}  \right] \,, \\
%	\kappaKinSixKinSix =& {\rm Tr}\left[ \left( \coef{6}{e^2 H^2 D}{} \right)^{2} + 2 \left(\coef{6}{\ell^2 H^2 D}{(1)}  \right)^{2} + N_{c} \left( \coef{6}{u^2 H^2 D}{} \right)^{2} + N_{c} \left( \coef{6}{d^2 H^2 D}{} \right)^{2}     \right. \nonumber \\ & \left. \qquad    + 2 N_{c} \left( \coef{6}{q^2 H^2 D}{(1)} \right)^{2} \right] \,, \\
%	\kappaKinSixKinSixSUTwo =& {\rm Tr}\left[ 2 \left( \coef{6}{\ell^2 H^2 D}{(3)} \right)^{2}  + 2 N_{c} \left( \coef{6}{q^2 H^2 D}{(3)} \right)^{2} \right] \,, \\
%	\kappaKinSixKinSixud =& {\rm Tr}\left[ 2 N_{c} \left(\coef{6}{udH^2 D}{}  \right) \left( \coef{6}{udH^2 D}{\dagger} \right) \right] \,.
%\end{align}
%
%For terms involving the two Yukawa couplings, we define 
%
%\begin{align}
%    \kappaYukawaFourYukawaSix =& {\rm Tr}\left[ Y_{e} \,\,\coef{6}{\ell e H^3}{} + N_c Y_{u} \,\,\coef{6}{quH^3}{} + N_c Y_{d} \,\,\coef{6}{qdH^3}{}  + {\rm h.c.} \right] \,, \\ 
%    \kappaYukawaFourYukawaEight =& {\rm Tr}\left[ Y_{e} \,\,\coef{8}{\ell e H^5}{} + N_c Y_{u} \,\,\coef{8}{quH^5}{} + N_c Y_{d} \,\,\coef{8}{qdH^5}{}  + {\rm h.c.} \right] \,, \\ 
    \kappaYukawaFourYukawaEightPrime =& {\rm Tr}\left[ Y^{\dagger}_{e} \,\,\coef{8}{\ell e H^5}{\dagger} + N_c Y_{u} \,\,\coef{8}{quH^5}{} + N_c Y^{\dagger}_{d} \,\,\coef{8}{qdH^5}{\dagger}  \right] \,, \\ 
%    \kappaYukawaSixYukawaSix =& {\rm Tr}\left[ \coef{6}{\ell e H^3}{} \,\,\coef{6}{\ell e H^3}{\dagger} + N_c \,\,\coef{6}{quH^3}{} \,\,\coef{6}{quH^3}{\dagger} + N_c \,\,\coef{6}{qdH^3}{} \,\,\coef{6}{qdH^3}{\dagger}  \right] \,, \\ 
%	\kappaYukawaKinSix =& {\rm Tr}\left[- Y_{e} Y^{\dagger}_{e} \coef{6}{e^2 H^2 D}{} + Y^{\dagger}_{e} Y_{e} \coef{6}{\ell^2 H^2 D}{(1)} - N_c Y_{d} Y^{\dagger}_{d} \coef{6}{d^2 H^2 D}{} + N_c Y^{\dagger}_{d} Y_{d} \coef{6}{q^2 H^2 D}{(1)}  \right.\nonumber \\ & \left. \qquad + N_c  Y_{u} Y^{\dagger}_{u}  \coef{6}{u^2 H^2 D}{} - N_c Y^{\dagger}_{u} Y_{u}  \coef{6}{q^2 H^2 D}{(1)} \right] \,, \\
 	\kappaYukawaKinEight =& {\rm Tr}\left[- Y_{e} Y^{\dagger}_{e} \coef{8}{e^2 H^4 D}{} + Y^{\dagger}_{e} Y_{e} \coef{8}{\ell^2 H^4 D}{(1)} - N_c Y_{d} Y^{\dagger}_{d} \coef{8}{d^2 H^4 D}{} + N_c Y^{\dagger}_{d} Y_{d} \coef{8}{q^2 H^4 D}{(1)}  \right.\nonumber \\ & \left. \qquad + N_c  Y_{u} Y^{\dagger}_{u}  \coef{8}{u^2 H^4 D}{} - N_c Y^{\dagger}_{u} Y_{u}  \coef{8}{q^2 H^4 D}{(1)} \right] \,, \\
 %%%%% Complete %%%%%%
% 	\kappaYukawaKinThreeSix =& {\rm Tr}\left[ Y^{\dagger}_{e} Y_{e}  \coef{6}{\ell^2 H^2 D}{(3)} + N_c \left( Y^{\dagger}_{d} Y_{d} + Y^{\dagger}_{u} Y_{u} \right) \coef{6}{q^2 H^2 D}{(3)} \right] \,, \\
   	\kappaYukawaKinThreeEight =& {\rm Tr}\left[ Y^{\dagger}_{e} Y_{e}  \coef{8}{\ell^2 H^4 D}{(3)} + N_c \left( Y^{\dagger}_{d} Y_{d} + Y^{\dagger}_{u} Y_{u} \right) \coef{8}{q^2 H^4 D}{(3)} \right] \,, \\
%    \kappaYukawaKinudSix =& {\rm Tr}\left[ - N_c Y_{d} Y^{\dagger}_{u} \coef{6}{udH^2 D}{}  - N_c Y_{u} Y^{\dagger}_{d} \coef{6}{udH^2 D}{\dagger} \right] \,, \\
%removed
    %\kappaYukawaKinudEight =& {\rm Tr}\left[ - N_c Y_{d} Y^{\dagger}_{u} \coef{8}{udH^4 D}{}  - N_c Y_{u} Y^{\dagger}_{d} \coef{8}{udH^4 D}{\dagger} \right] \,, \\
    \kappaYukawaKinTwoEight =& {\rm Tr}\left[ Y^{\dagger}_{e} Y_{e}  \coef{8}{\ell^2 H^4 D}{(2)} + N_c \left( Y^{\dagger}_{d} Y_{d} + Y^{\dagger}_{u} Y_{u} \right) \coef{8}{q^2 H^4 D}{(2)} \right] \,.
\end{align}
%

%%%%%%%%%%%%%%%%%%%%%%%%%%%%%%%%%%%%%%%%%%%%%%%%
\subsection{Dimension 6}
%%%%%%%%%%%%%%%%%%%%%%%%%%%%%%%%%%%%%%%%%%%%%%%%

The RGEs for the dimension-six coefficients in the SMEFT Lagrangian that arise from dimension-eight operators are listed below. The dimension-eight contributions are all of order $m_H^2/M^4$ in the SMEFT power counting. The dimension-six are given in refs.~\cite{Jenkins:2013zja,Jenkins:2013wua,Alonso:2013hga}, and some results from products of dimension-six operators are given in ref.~\cite{Bakshi:2024wzz}.

%%%%%%%%%%%%%%%%%%%%%%%%%%%%%%%%%%%%%%%%%%%%%%%%
\subsubsection{$\psi^2 H^3$}
%%%%%%%%%%%%%%%%%%%%%%%%%%%%%%%%%%%%%%%%%%%%%%%%

The RGEs for the $\psi^2 H^3$ couplings are
\begin{align}
    \dcoef{6}{\underset{\bar p r}{\ell e H^3}}{} =  -2v^2\lambda &\Bigg[~
    8 \coef{8}{\underset{\bar p r}{\ell e H^5}}{} +  \left( 3\coef{8}{H^6 D^2}{(1)} -2\coef{8}{H^6 D^2}{(2)} \right)Y_{\underset{\bar p r}{e}}^{\dagger}  
    \nn &
    + \left[  \left( \coef{8}{\ell^2 H^4 D}{(1)} + \coef{8}{\ell^2 H^4 D}{(2)} + 3\coef{8}{\ell^2 H^4 D}{(3)} + 4i\coef{8}{\ell^2 H^4 D}{(4)}  \right) Y^{\dagger}_{e} \right]_{\bar p r}
    \nn & 
    - \left[  Y_{e}  \coef{8}{e^2 H^4 D}{}   \right]_{\bar p r} 
    \Bigg]
    \,,
\end{align}
\begin{align}
    \dcoef{6}{\underset{\bar p r}{q u H^3}}{} =  -2v^2\lambda &\Bigg[
    ~ 8 \coef{8}{\underset{\bar p r}{q u H^5}}{} +  \left( 3\coef{8}{H^6 D^2}{(1)} -2\coef{8}{H^6 D^2}{(2)} \right)Y_{\underset{\bar p r}{u}}^{\dagger} 
    \nn &
    + \left[   \left( -\coef{8}{q^2 H^4 D}{(1)} + \coef{8}{q^2 H^4 D}{(2)} + 3\coef{8}{q^2 H^4 D}{(3)} - 4i\coef{8}{q^2 H^4 D}{(4)}  \right) Y^{\dagger}_{u}  \right]_{\bar p r}
    \nn & 
    + \left[   Y_{u} \coef{8}{u^2 H^4 D}{}  - Y_{d}\coef{8}{ud H^4 D}{\dagger} \right]_{\bar p r}
    \Bigg]
    \,,
\end{align}
\begin{align}
    \dcoef{6}{\underset{\bar p r}{q d H^3}}{} =  -2v^2\lambda &\Bigg[
    ~ 8 \coef{8}{\underset{\bar p r}{q d H^5}}{} + \left( 3\coef{8}{H^6 D^2}{(1)} -2\coef{8}{H^6 D^2}{(2)} \right)Y_{\underset{\bar p r}{d}}^{\dagger}    
    \nn &
    + \left[  \left( \coef{8}{q^2 H^4 D}{(1)} + \coef{8}{q^2 H^4 D}{(2)} + 3\coef{8}{q^2 H^4 D}{(3)} + 4i\coef{8}{q^2 H^4 D}{(4)}  \right) Y^{\dagger}_{d}  \right]_{\bar p r}
    \nn & 
    + \left[   - Y_{d}\coef{8}{d^2 H^4 D}{} - Y_{u}\coef{8}{ud H^4 D}{}   \right]_{\bar p r}
    \Bigg]
    \,.
\end{align}
%%

%%%%%%%%%%%%%%%%%%%%%%%%%%%%%%%%%%%%%%%%%%%%%%%%
\subsubsection{$\psi^2 H^2 D$}
%%%%%%%%%%%%%%%%%%%%%%%%%%%%%%%%%%%%%%%%%%%%%%%%

The RGEs for the $\psi^2 H^2 D$ couplings are
\begin{align}
    \dcoef{6}{\underset{\bar p r}{e^2 H^2 D}}{} =& 
    \left( - 6 v^2 \lambda \right) \coef{8}{\underset{\bar p r}{e^2 H^4 D}}{}  
     \,,
\\
    \dcoef{6}{\underset{\bar p r}{\ell^2 H^2 D}}{(1)} =& 
    \left( - 6 v^2 \lambda \right) \coef{8}{\underset{\bar p r}{\ell^2 H^4 D}}{(1)}  
     \,,
\\
    \dcoef{6}{\underset{\bar p r}{\ell^2 H^2 D}}{(3)} =& 
    \left( -2 v^2\lambda \right) \left( \coef{8}{\underset{\bar p r}{\ell^2 H^4 D}}{(2)} + 3 \coef{8}{\underset{\bar p r}{\ell^2 H^4 D}}{(3)}    \right)
      \,,
\\
    \dcoef{6}{\underset{\bar p r}{u^2 H^2 D}}{} =& 
    \left( - 6 v^2 \lambda \right) \coef{8}{\underset{\bar p r}{u^2 H^4 D}}{}  
      \,,
\\
    \dcoef{6}{\underset{\bar p r}{d^2 H^2 D}}{} =& 
    \left( - 6 v^2 \lambda \right) \coef{8}{\underset{\bar p r}{d^2 H^4 D}}{}  
      \,,
\\
    \dcoef{6}{\underset{\bar p r}{q^2 H^2 D}}{(1)} =& 
    \left( - 6 v^2 \lambda \right) \coef{8}{\underset{\bar p r}{q^2 H^4 D}}{(1)}  
      \,,
\\
    \dcoef{6}{\underset{\bar p r}{q^2 H^2 D}}{(3)} =&
    \left( -2 v^2\lambda \right) \left( \coef{8}{\underset{\bar p r}{q^2 H^4 D}}{(2)} +3 \coef{8}{\underset{\bar p r}{q^2 H^4 D}}{(3)}  \right)
     \,. 
\\
    \dcoef{6}{\underset{\bar p r}{ud H^2 D}}{} =& \left(-6 v^2 \lambda\right) \coef{8}{\underset{\bar p r}{ud H^4 D}}{}  
    \,. 
\end{align}
%

%%%%%%%%%%%%%%%%%%%%%%%%%%%%%%%%%%%%%%%%%%%%
\subsection{Dimension 8}
%%%%%%%%%%%%%%%%%%%%%%%%%%%%%%%%%%%%%%%%%%%%

The RGEs for the dimension-eight coefficients in the SMEFT Lagrangian that arise from dimension-eight operators are listed below. Some results from products of dimension-six operators are given in ref.~\cite{Bakshi:2024wzz}.

%%%%%%%%%%%%%%%%%%%%%%%%%%%%%%%%%%%%%%%%%%%%%%%%
\subsubsection{$\psi^2 H^5$}
%%%%%%%%%%%%%%%%%%%%%%%%%%%%%%%%%%%%%%%%%%%%%%%%

The RGEs for the $\psi^2 H^5$ couplings are
\begin{align}
    \dcoef{8}{\underset{\bar p r}{\ell e H^5}}{}
    =&  
    \left( 72 \lambda  + 5 \gamma_{H}^{(Y)} + \left(-4 + y_e^2 +  y_\ell^2    - 8 y_\ell y_e \right) g^2_1  -\frac{45}{4} g_2^2  \right) \coef{8}{\underset{\bar p r}{\ell e H^5}}{} 
    \nn &
    +\left[ \gamma^{(Y)}_{\ell} \coef{8}{\underset{}{\ell e H^5}}{} +  \coef{8}{\underset{}{\ell e H^5}}{} \gamma^{(Y)}_{e}  \right]_{\bar pr} \nn &
    -  \frac{1}{2} \left((   g_1^2 + 3 g_2^2 )\coef{8}{H^6D^2}{(1)} -( 24 \lambda - g_1^2 + 3 g_2^2 )\coef{8}{H^6D^2}{(2)} \right) Y^{\dagger}_{\underset{\bar p r}{e}}  
    \nn &
    +  8 \left(  g_1^2 \coef{8}{B^2H^4}{(1)} +   g_1 g_2 \coef{8}{WBH^4}{(1)} +  g_2^2 (3\coef{8}{W^2H^4}{(1)} + \coef{8}{W^2H^4}{(3)}) \right) Y^{\dagger}_{\underset{\bar p r}{e}}
    \nn &
    + Y^{\dagger}_{\underset{\bar p r}{e}} \dimEightBosonLoopYuk  
    - \frac{1}{3} g_2^2 Y^{\dagger}_{\underset{\bar p r}{e}} \left( \kappaPauliKinEight +\kappaPauliKinEightTwo \right)
    %\nn  & 
    + Y^{\dagger}_{\underset{\bar p r}{e}} \left(2(\kappaYukawaFourYukawaEightPrime)^\dagger  + \kappaYukawaKinEight + \kappaYukawaKinThreeEight  + \kappaYukawaKinTwoEight  \right)
    \nn &
    +  \left( \frac{3}{2} \coef{8}{H^6D^2}{(1)} + \frac{1}{2} \coef{8}{H^6D^2}{(2)} \right) \left[Y^{\dagger}_{e} Y_{e} Y^{\dagger}_{e}\right]_{\bar p r}  
    \nn & 
    +  \left( 4 \lambda + g_1^2 + g_2^2 \right) \left[ \coef{8}{\ell^2 H^4 D}{(1)} Y^{\dagger}_{e} + \coef{8}{\ell^2 H^4 D}{(2)} Y^{\dagger}_{e}- Y_{e} \coef{8}{e^2 H^4 D}{(1)}    \right]_{\bar p r}
    \nn & 
    +  \left( 12 \lambda + g_1^2 + 3 g_2^2 \right) \left[ \coef{8}{\ell^2 H^4 D}{(3)} Y^{\dagger}_{e}  \right]_{\bar p r}
    \nn & 
    + 2   \left( 4 \lambda +  g_2^2 \right) \left[ i \coef{8}{\ell^2 H^4 D}{(4)} Y^{\dagger}_{e}  \right]_{\bar p r}    
    \nn & 
    +  2 \left[ \left( \coef{8}{\ell^2 H^4 D}{(1)} + \coef{8}{\ell^2 H^4 D}{(2)} - i \coef{8}{\ell^2 H^4 D}{(4)} \right) Y^{\dagger}_{e} Y_{e} Y^{\dagger}_{e}  \right]_{\bar pr} 
    \nn &
     - 2 \left[  Y^{\dagger}_{e} Y_{e} Y^{\dagger}_{e}  \coef{8}{e^2 H^4 D}{}  \right]_{\bar pr}  
    \,,
\end{align}
\begin{align}
    \dcoef{8}{\underset{\bar p r}{q u H^5}}{} 
    =&  \left( 72 \lambda  + 5 \gamma_{H}^{(Y)} + \left(-4 + y_u^2 +  y_q^2    - 8 y_q y_u \right) g^2_1  -\frac{45}{4} g_2^2  \right)  \coef{8}{\underset{\bar p r}{q u H^5}}{} \nn &
    +\left[ \gamma^{(Y)}_{q} \coef{8}{\underset{}{q u H^5}}{} +  \coef{8}{\underset{}{q u H^5}}{} \gamma^{(Y)}_{u}  \right]_{\bar pr} \nn &
    -   \frac{1}{2} \left((   g_1^2 + 3 g_2^2 )\coef{8}{H^6D^2}{(1)} -( 24 \lambda - g_1^2 + 3 g_2^2 )\coef{8}{H^6D^2}{(2)} \right) Y^{\dagger}_{\underset{\bar p r}{u}}  
    \nn &
    +  8\left(  g_1^2 \coef{8}{B^2H^4}{(1)} +   g_1 g_2 \coef{8}{WBH^4}{(1)} +  g_2^2 (3\coef{8}{W^2H^4}{(1)} +  \coef{8}{W^2H^4}{(3)}) \right) Y^{\dagger}_{\underset{\bar p r}{u}}
    \nn &
    + Y^{\dagger}_{\underset{\bar p r}{u}} \dimEightBosonLoopYuk   
    - \frac{1}{3} g_2^2 Y^{\dagger}_{\underset{\bar p r}{u}} \left( \kappaPauliKinEight +\kappaPauliKinEightTwo \right)
    %\nn  & 
    + Y^{\dagger}_{\underset{\bar p r}{u}} \left(2\kappaYukawaFourYukawaEightPrime + \kappaYukawaKinEight + \kappaYukawaKinThreeEight  + \kappaYukawaKinTwoEight \right) 
    \nn &
    +  \left( \frac{3}{2} \coef{8}{H^6D^2}{(1)} + \frac{1}{2} \coef{8}{H^6D^2}{(2)} \right) \left[Y^{\dagger}_{u} Y_{u} Y^{\dagger}_{u}\right]_{\bar p r}  
    \nn & 
    +  \left( 4 \lambda + g_1^2 + g_2^2 \right) \left[ -\coef{8}{q^2 H^4 D}{(1)} Y^{\dagger}_{u} + \coef{8}{\ell^2 H^4 D}{(2)} Y^{\dagger}_{u} + Y_{u} \coef{8}{u^2 H^4 D}{}    \right]_{\bar p r}
    \nn & 
    +  \left( 12 \lambda + g_1^2 + 3 g_2^2 \right) \left[ \coef{8}{q^2 H^4 D}{(3)} Y^{\dagger}_{u}  \right]_{\bar p r}
    \nn & 
    -  2 \left( 4 \lambda +  g_2^2 \right) \left[ i \coef{8}{q^2 H^4 D}{(4)} Y^{\dagger}_{u} \right]_{\bar p r} 
    \nn & 
    + \left( 4 \lambda  + g_2^2 \right) \left[   Y_{d} \coef{8}{ud H^4 D}{\dagger}   \right]_{\bar p r}
    \nn &
     + 2 \left[ \left( -\coef{8}{q^2 H^4 D}{(1)} + \coef{8}{q^2 H^4 D}{(2)} + i \coef{8}{q^2 H^4 D}{(4)} \right) Y^{\dagger}_{u} Y_{u} Y^{\dagger}_{u}  \right]_{\bar pr}  
    \nn &
     + 6 \left[ \left( \coef{8}{q^2 H^4 D}{(3)} - i \coef{8}{q^2 H^4 D}{(4)} \right) Y^{\dagger}_{d} Y_{d} Y^{\dagger}_{u}  \right]_{\bar pr} 
    \nn &
     + 2 \left[  Y^{\dagger}_{u} Y_{u} Y^{\dagger}_{u}  \coef{8}{u^2 H^4 D}{}  \right]_{\bar pr} - 2 \left[  Y^{\dagger}_{d} Y_{d} Y^{\dagger}_{d}  \coef{8}{ud H^4 D}{\dagger}  \right]_{\bar pr} 
    \,,
\end{align}
\begin{align}
    \dcoef{8}{\underset{\bar p r}{q d H^5}}{} 
    =&  
    \left( 72 \lambda  + 5 \gamma_{H}^{(Y)} + \left(-4 + y_d^2 +  y_q^2    - 8 y_q y_d \right) g^2_1  -\frac{45}{4} g_2^2  \right) \coef{8}{\underset{\bar p r}{q d H^5}}{} \nn &
    +\left[ \gamma^{(Y)}_{q} \coef{8}{\underset{}{q d H^5}}{} +  \coef{8}{\underset{}{q d H^5}}{} \gamma^{(Y)}_{d}  \right]_{\bar pr} \nn &
    -  \frac{1}{2} \left((   g_1^2 + 3 g_2^2 )\coef{8}{H^6D^2}{(1)} -( 24 \lambda - g_1^2 + 3 g_2^2 )\coef{8}{H^6D^2}{(2)} \right) Y^{\dagger}_{\underset{\bar p r}{d}}  
    \nn &
    + 8 \left(  g_1^2 \coef{8}{B^2H^4}{(1)} +   g_1 g_2 \coef{8}{WBH^4}{(1)} +  g_2^2 (3\coef{8}{W^2H^4}{(1)} +  \coef{8}{W^2H^4}{(3)} )\right) Y^{\dagger}_{\underset{\bar p r}{d}}
    \nn &
    + Y^{\dagger}_{\underset{\bar p r}{d}} \dimEightBosonLoopYuk - \frac{1}{3} g_2^2 Y^{\dagger}_{\underset{\bar p r}{d}} \left( \kappaPauliKinEight +\kappaPauliKinEightTwo \right)
    %\nn  & 
    + Y^{\dagger}_{\underset{\bar p r}{d}} \left(2(\kappaYukawaFourYukawaEightPrime)^{\dagger} + \kappaYukawaKinEight + \kappaYukawaKinThreeEight  + \kappaYukawaKinTwoEight \right) 
    \nn &
    +  \left( \frac{3}{2} \coef{8}{H^6D^2}{(1)} + \frac{1}{2} \coef{8}{H^6D^2}{(2)} \right) \left[Y^{\dagger}_{d} Y_{d} Y^{\dagger}_{d}\right]_{\bar p r}  
    \nn & 
    +  \left( 4 \lambda + g_1^2 + g_2^2 \right) \left[\coef{8}{q^2 H^4 D}{(1)} Y^{\dagger}_{d} + \coef{8}{q^2 H^4 D}{(2)} Y^{\dagger}_{d}- Y_{d} \coef{8}{d^2 H^4 D}{(1)}    \right]_{\bar p r}
    \nn & 
    +  \left( 12 \lambda + g_1^2 + 3 g_2^2 \right) \left[ \coef{8}{q^2 H^4 D}{(3)} Y^{\dagger}_{d}   \right]_{\bar p r}
    \nn & 
    +  2\left( 4 \lambda +  g_2^2 \right) \left[ i\coef{8}{q^2 H^4 D}{(4)} Y^{\dagger}_{d} \right]_{\bar p r} 
    \nn & 
    -  \left( 4 \lambda  + g_2^2 \right) \left[   Y_{u} \coef{8}{ud H^4 D}{}   \right]_{\bar p r}
    \nn &
     + 2 \left[ \left( \coef{8}{q^2 H^4 D}{(1)} + \coef{8}{q^2 H^4 D}{(2)} - i \coef{8}{q^2 H^4 D}{(4)} \right) Y^{\dagger}_{d} Y_{d} Y^{\dagger}_{d}  \right]_{\bar pr}  
    \nn &
     + 6 \left[ \left( \coef{8}{q^2 H^4 D}{(3)} + i \coef{8}{q^2 H^4 D}{(4)}\right) Y^{\dagger}_{u} Y_{u} Y^{\dagger}_{d}  \right]_{\bar pr}  
    \nn &
    - 2 \left[  Y^{\dagger}_{d} Y_{d} Y^{\dagger}_{d}  \coef{8}{d^2 H^4 D}{}  \right]_{\bar pr} - 2 \left[  Y^{\dagger}_{u} Y_{u} Y^{\dagger}_{u}  \coef{8}{ud H^4 D}{}  \right]_{\bar pr}
    \,,
\end{align}
%%

%%%%%%%%%%%%%%%%%%%%%%%%%%%%%%%%%%%%%%%%%%%%%%%%
\subsubsection{$\psi^2 H^4 D$}
%%%%%%%%%%%%%%%%%%%%%%%%%%%%%%%%%%%%%%%%%%%%%%%%

The RGEs for the $\psi^2 H^4 D$ couplings are
\begin{align}
    \dcoef{8}{\underset{\bar p r}{e^2 H^4 D}}{} 
    =& 
    \left( 28 \lambda  + 4 \gamma_{H}^{(Y)} +\frac{1}{2} g_1^2  -\frac{17}{6}  g_2^2    \right) \coef{8}{\underset{\bar p r}{e^2 H^4 D}}{}
    \nn &
    +\left[ \gamma^{(Y)}_{e} \coef{8}{\underset{}{e^2 H^4 D}}{} +  \coef{8}{\underset{}{e^2 H^4 D}}{} \gamma^{(Y)}_{e}  \right]_{\bar pr} \nn &
    +\left(   8 y_e g_1^2 \coef{8}{B^2H^4}{(1)} + 8 y_e g_1 g_2 \coef{8}{WBH^4}{(1)} + y_{e} \dimEightBosonLoopOne 
    +\frac{4}{3} y_{e} g_{1}^2 (\kappaHypKinEight) \right) \delta_{\bar pr}   
    \nn &
    +  \left[Y_{e}Y_{e}^{\dagger}\right]_{\bar pr} \left( \coef{8}{H^6D^2}{(1)} + 2\coef{8}{H^6D^2}{(2)} \right)  
    \nn &
     - 2 \left[Y_{e} \coef{8}{\ell^2 H^4 D}{(1)} Y^{\dagger}_{e}\right]_{\bar pr} + 6 \left[\coef{8}{e^2 H^4 D}{} Y_{e} Y^{\dagger}_{e} + Y_{e} Y^{\dagger}_{e} \coef{8}{e^2 H^4 D}{} \right]_{\bar pr} 
    \,,
\end{align}
\begin{align}
    \dcoef{8}{\underset{\bar p r}{\ell^2 H^4 D}}{(1)} 
    =& 
    \left( 28 \lambda + 4  \gamma_{H}^{(Y)} + \frac{1}{2} g_1^2 - \frac{17}{6} g_2^2  \right) \coef{8}{\underset{\bar p r}{\ell^2 H^4 D}}{(1)}
    \nn &
    +\left[ \gamma^{(Y)}_{\ell} \coef{8}{\underset{}{\ell^2 H^4 D}}{(1)} +  \coef{8}{\underset{}{\ell^2 H^4 D}}{(1)} \gamma^{(Y)}_{\ell}  \right]_{\bar pr} 
    \nn &
    +  \left( 8 y_\ell g_1^2 \coef{8}{B^2H^4}{(1)} + 8 y_\ell g_1 g_2 \coef{8}{WBH^4}{(1)} + y_{\ell} \dimEightBosonLoopOne 
    + \frac{4}{3} y_{\ell} g_{1}^2 (\kappaHypKinEight) \right)\delta_{\bar pr}  
    \nn  & 
     -  \frac{1}{2} \left[Y_{e}^{\dagger}Y_{e}\right]_{\bar pr}\left( \coef{8}{H^6D^2}{(1)} + 2\coef{8}{H^6D^2}{(2)} \right)  
    \nn &
     -  \left[Y^{\dagger}_{e} \coef{8}{e^2 H^4 D}{} Y_{e}\right]_{\bar pr} 
     + 3 \left[ \left( \coef{8}{\ell^2 H^4 D}{(1)} + 2\coef{8}{\ell^2 H^4 D}{(2)} + 2\coef{8}{\ell^2 H^4 D}{(3)} \right) Y^{\dagger}_{e}Y_{e}\right]_{\bar pr}\nonumber \\ &  
    + 3\left[ Y^{\dagger}_{e}Y_{e}\left( \coef{8}{\ell^2 H^4 D}{(1)} + 2\coef{8}{\ell^2 H^4 D}{(2)} + 2\coef{8}{\ell^2 H^4 D}{(3)}\right) \right]_{\bar pr} 
    \,,
\end{align}
\begin{align}
    \dcoef{8}{\underset{\bar p r}{\ell^2 H^4 D}}{(2)} 
    =& 
    \left( 16 \lambda  +4  \gamma_{H}^{(Y)}    - g_1^2   -2 g_2^2   \right) \coef{8}{\underset{\bar p r}{\ell^2 H^4 D}}{(2)}
    \nn &    
    +\left[ \gamma^{(Y)}_{\ell} \coef{8}{\underset{}{\ell^2 H^4 D}}{(2)} +  \coef{8}{\underset{}{\ell^2 H^4 D}}{(2)} \gamma^{(Y)}_{\ell}  \right]_{\bar pr} \nn &
    + \frac{5}{3}  g_1^2  \coef{8}{\underset{\bar p r}{\ell^2 H^4 D}}{(3)} 
    \nn &
      +  \left( g_1 g_2 \coef{8}{WBH^4}{(1)} + 10 g_2^2 \coef{8}{W^2H^4}{(3)} + \dimEightBosonLoopTwo 
    + \frac{1}{3} g_{2}^2 \left( 2\kappaPauliKinEightTwo\right) \right)\delta_{\bar pr}  
    \nn  & 
     - \frac{1}{2}\left[Y_{e}^{\dagger}Y_{e}\right]_{\bar pr}  \coef{8}{H^6D^2}{(2)}  
    \nn &
     + \frac{1}{2}\left[ \left( 3\coef{8}{\ell^2 H^4 D}{(1)} + 4\coef{8}{\ell^2 H^4 D}{(2)}  +6i\coef{8}{\ell^2 H^4 D}{(4)}\right) Y^{\dagger}_{e}Y_{e}  + \textrm{h.c.}  \right]_{\bar pr} 
    \,,
\end{align}
\begin{align}
    \dcoef{8}{\underset{\bar p r}{\ell^2 H^4 D}}{(3)} 
    =& 
    \left(  28  \lambda  +4  \gamma_{H}^{(Y)} - \frac{3}{2} g_1^2  - \frac{9}{2} g_2^2  \right) \coef{8}{\underset{\bar p r}{\ell^2 H^4 D}}{(3)}
    \nn &
    +\left[ \gamma^{(Y)}_{\ell} \coef{8}{\underset{}{\ell^2 H^4 D}}{(3)} +  \coef{8}{\underset{}{\ell^2 H^4 D}}{(3)} \gamma^{(Y)}_{\ell}  \right]_{\bar pr} 
    \nn &
    +\left( 4 \lambda  + \frac{3}{2} g_1^2  - \frac{5}{6}  g_2^2 \right) \coef{8}{\underset{\bar p r}{\ell^2 H^4 D}}{(2)} 
    \nn &
    +\left(   g_1 g_2 \coef{8}{WBH^4}{(1)} + 4 g_2^2 \coef{8}{W^2H^4}{(1)} - 2 g_2^2 \coef{8}{W^2H^4}{(3)} + \dimEightBosonLoopThree 
    +\frac{1}{3} g_{2}^2 \left( 2\kappaPauliKinEight\right) \right)\delta_{\bar pr} 
    \nn  & 
     - \frac{1}{2}\left[Y_{e}^{\dagger}Y_{e}\right]_{\bar pr} \left( \coef{8}{H^6D^2}{(1)} - \coef{8}{H^6D^2}{(2)} \right) 
    \nn &
     + \frac{1}{2}\left[ \left( 3\coef{8}{\ell^2 H^4 D}{(1)}  + 4\coef{8}{\ell^2 H^4 D}{(3)} -6i\coef{8}{\ell^2 H^4 D}{(4)}\right) Y^{\dagger}_{e}Y_{e} + \textrm{h.c.} \right]_{\bar pr} 
    \,,
\end{align}
\begin{align}
    \dcoef{8}{\underset{\bar p r}{\ell^2 H^4 D}}{(4)} 
    =& 
    \left(24\lambda + 4 \gamma_{H}^{(Y)} - 3 g_1^2 - \frac{17}{3} g_2^2 \right)\coef{8}{\underset{\bar p r}{\ell^2 H^4 D}}{(4)}
    \nn &
    +\left[ \gamma^{(Y)}_{\ell} \coef{8}{\underset{}{\ell^2 H^4 D}}{(4)} +  \coef{8}{\underset{}{\ell^2 H^4 D}}{(4)} \gamma^{(Y)}_{\ell}  \right]_{\bar pr} \nn &
    + \frac{1}{3} g_{2}^2 \left( 2\kappaPauliKinEightFour\right)\delta_{\bar pr}
    \nn &
    + \frac{1}{2} \left[ \left( - 3i\coef{8}{\ell^2 H^4 D}{(2)} +3i \coef{8}{\ell^2 H^4 D}{(3)} + 4\coef{8}{\ell^2 H^4 D}{(4)} \right) Y^{\dagger}_{e}Y_{e} + \textrm{h.c.} \right]_{\bar pr}
    \,,
\end{align}
\begin{align}
    \dcoef{8}{\underset{\bar p r}{u^2 H^4 D}}{} 
    =& 
    \left( 28 \lambda + 4 \gamma_{H}^{(Y)} + \frac{1}{2} g_1^2 - \frac{17}{6} g_2^2 \right) \coef{8}{\underset{\bar p r}{u^2 H^4 D}}{}
    \nn &
    +\left[ \gamma^{(Y)}_{u} \coef{8}{\underset{}{u^2 H^4 D}}{} +  \coef{8}{\underset{}{u^2 H^4 D}}{} \gamma^{(Y)}_{u}  \right]_{\bar pr} 
    \nn &
     + \left( 8 y_u g_1^2 \coef{8}{B^2H^4}{(1)} + 8 y_u g_1 g_2 \coef{8}{WBH^4}{(1)}  + y_{u} \dimEightBosonLoopOne 
    + \frac{4}{3} y_{u} g_{1}^2 (\kappaHypKinEight) \right) \delta_{\bar pr}  
    \nn  & 
     -\left[Y_{u}Y_{u}^{\dagger}\right]_{\bar pr} \left( \coef{8}{H^6D^2}{(1)} + 2\coef{8}{H^6D^2}{(2)} \right)
    \nn &
      + \left[-  Y_{u} \coef{8}{q^2 H^4 D}{(1)} Y^{\dagger}_{u} +  6 \coef{8}{u^2 H^4 D}{} Y_{u} Y^{\dagger}_{u} + \coef{8}{ud H^4 D}{} Y_{d} Y^{\dagger}_{u} + \textrm{h.c.} \right]_{\bar pr} 
    \,,
\end{align}
\begin{align}
    \dcoef{8}{\underset{\bar p r}{d^2 H^4 D}}{} 
    =& 
    \left( 28 \lambda + 4 \gamma_{H}^{(Y)} + \frac{1}{2} g_1^2 - \frac{17}{6} g_2^2  \right) \coef{8}{\underset{\bar p r}{d^2 H^4 D}}{}
    \nn &
    +\left[ \gamma^{(Y)}_{d} \coef{8}{\underset{}{d^2 H^4 D}}{} +  \coef{8}{\underset{}{d^2 H^4 D}}{} \gamma^{(Y)}_{d}  \right]_{\bar pr} 
    \nn &
    +\left( 8 y_d g_1^2 \coef{8}{B^2H^4}{(1)} + 8 y_d g_1 g_2 \coef{8}{WBH^4}{(1)}  + y_{d} \dimEightBosonLoopOne 
    + \frac{4}{3} y_{d} g_{1}^2 (\kappaHypKinEight)\right) \delta_{\bar pr} 
    \nn  & 
    + \left[ Y_{d}Y_{d}^{\dagger}\right]_{\bar pr} \left( \coef{8}{H^6D^2}{(1)} + 2\coef{8}{H^6D^2}{(2)} \right) 
    \nn &
     + \left[-Y_{d} \coef{8}{q^2 H^4 D}{(1)} Y^{\dagger}_{d} + 6 \coef{8}{d^2 H^4 D}{} Y_{d} Y^{\dagger}_{d} -  \coef{8}{ud H^4 D}{\dagger} Y_{u} Y^{\dagger}_{d}     + \textrm{h.c.}  \right]_{\bar pr} 
    \,,
\end{align}
\begin{align}
    \dcoef{8}{\underset{\bar p r}{q^2 H^4 D}}{(1)}
    =& 
    \left[ 28 \lambda + 4 \gamma_{H}^{(Y)} + \frac{1}{2} g_1^2 - \frac{17}{6} g_2^2   \right] \coef{8}{\underset{\bar p r}{q^2 H^4 D}}{(1)}
    \nn &
    +\left[ \gamma^{(Y)}_{q} \coef{8}{\underset{}{q^2 H^4 D}}{(1)} +  \coef{8}{\underset{}{q^2 H^4 D}}{(1)} \gamma^{(Y)}_{q}  \right]_{\bar pr} 
    \nn &
    +\left( 8 y_q g_1^2 \coef{8}{B^2H^4}{(1)} + 8 y_q g_1 g_2 \coef{8}{WBH^4}{(1)} + y_{q} \dimEightBosonLoopOne 
    + \frac{4}{3} y_{q} g_{1}^2 (\kappaHypKinEight) \right)\delta_{\bar pr}
    \nn  & 
    + \frac{1}{2} \left[  Y_{u}^{\dagger}Y_{u} - Y_{d}^{\dagger}Y_{d}\right]_{\bar pr} \left( \coef{8}{H^6D^2}{(1)} + 2\coef{8}{H^6D^2}{(2)} \right)  
    \nn &
    + \bigg[ - \frac{1}{2} Y^{\dagger}_{u} \coef{8}{u^2 H^4 D}{} Y^{\dagger}_{u} - \frac{1}{2} Y^{\dagger}_{d} \coef{8}{d^2 H^4 D}{} Y^{\dagger}_{d}  
    \nonumber \\ &  \qquad +  3 \left( \coef{8}{q^2 H^4 D}{(1)} + 2\coef{8}{q^2 H^4 D}{(2)} + 2\coef{8}{q^2 H^4 D}{(3)}  \right)  Y^{\dagger}_{d}Y_{d}\nonumber \\ &  \qquad
    +  3 \left( \coef{8}{q^2 H^4 D}{(1)} - 2\coef{8}{q^2 H^4 D}{(2)} - 2\coef{8}{q^2 H^4 D}{(3)}\right) Y^{\dagger}_{u}Y_{u}  + \textrm{h.c.} \bigg]_{\bar pr}
    \,,
\end{align}
\begin{align}
    \dcoef{8}{\underset{\bar p r}{q^2 H^4 D}}{(2)} 
    =&  
    \left( 16 \lambda + 4 \gamma_{H}^{(Y)} - g_1^2 - 2 g_2^2 \right) \coef{8}{\underset{\bar p r}{q^2 H^4 D}}{(2)}
    \nn &
    +\left[ \gamma^{(Y)}_{q} \coef{8}{\underset{}{q^2 H^4 D}}{(2)} +  \coef{8}{\underset{}{q^2 H^4 D}}{(2)} \gamma^{(Y)}_{q}  \right]_{\bar pr} 
    \nn &
    +  \frac{5}{3} g_1^2  \coef{8}{\underset{\bar p r}{q^2 H^4 D}}{(3)} 
    \nn &
     +  \left( g_1 g_2 \coef{8}{WBH^4}{(1)} + 10 g_2^2 \coef{8}{W^2H^4}{(3)} + \dimEightBosonLoopTwo 
    + \frac{1}{3} g_{2}^2 \left( 2\kappaPauliKinEightTwo \right) \right)\delta_{\bar pr}
    \nn  & 
      - \frac{1}{2}\left[Y_{u}^{\dagger}Y_{u} + Y_{d}^{\dagger}Y_{d}\right]_{\bar pr}  \coef{8}{H^6D^2}{(2)} 
    \nn &
     + \frac{1}{2} \bigg[ \left( 3\coef{8}{q^2 H^4 D}{(1)} + 4\coef{8}{q^2 H^4 D}{(2)}  + 6i\coef{8}{q^2 H^4 D}{(4)}\right) Y^{\dagger}_{d}Y_{d}\nonumber \\ & \qquad
    +  \left( -3\coef{8}{q^2 H^4 D}{(1)} + 4\coef{8}{q^2 H^4 D}{(2)}  -6i\coef{8}{q^2 H^4 D}{(4)}\right) Y^{\dagger}_{u}Y_{u}  + \textrm{h.c.} \bigg]_{\bar pr}
    \,,
\end{align}
\begin{align}
    \dcoef{8}{\underset{\bar p r}{q^2 H^4 D}}{(3)} 
    =& 
    \left( 28 \lambda + 4 \gamma_{H}^{(Y)} - \frac{3}{2} g_1^2 - \frac{9}{2} g_2^2   \right) \coef{8}{\underset{\bar p r}{q^2 H^4 D}}{(3)}
    \nn &
    +\left[ \gamma^{(Y)}_{q} \coef{8}{\underset{}{q^2 H^4 D}}{(3)} +  \coef{8}{\underset{}{q^2 H^4 D}}{(3)} \gamma^{(Y)}_{q}  \right]_{\bar pr} \nn &
    +\left( 4 \lambda + \frac{3}{2}  g_1^2 - \frac{5}{6} g_2^2 \right) \coef{8}{\underset{\bar p r}{q^2 H^4 D}}{(2)} 
    \nn &
    +\left(  g_1 g_2 \coef{8}{WBH^4}{(1)} + 4 g_2^2 \coef{8}{W^2H^4}{(1)} - 2 g_2^2 \coef{8}{W^2H^4}{(3)} + \dimEightBosonLoopThree 
    +\frac{1}{3} g_{2}^2 \left( 2\kappaPauliKinEight \right)  \right) \delta_{\bar pr}
    \nn  & 
     - \frac{1}{2}\left[Y_{u}^{\dagger}Y_{u} + Y_{d}^{\dagger}Y_{d}\right]_{\bar pr} \left( \coef{8}{H^6D^2}{(1)} - \coef{8}{H^6D^2}{(2)} \right) 
    \nn &
    + \frac{1}{2}\left[ \left( 3\coef{8}{q^2 H^4 D}{(1)} + 4\coef{8}{q^2 H^4 D}{(3)} -6i\coef{8}{q^2 H^4 D}{(4)}\right) Y^{\dagger}_{d}Y_{d} \right.\nonumber \\ & \left. \qquad
    +  \left( -3\coef{8}{q^2 H^4 D}{(1)}  + 4\coef{8}{q^2 H^4 D}{(3)} +6i\coef{8}{q^2 H^4 D}{(4)}\right) Y^{\dagger}_{u}Y_{u} + \textrm{h.c.} \right]_{\bar pr}
    \,,
\end{align}
\begin{align}
    \dcoef{8}{\underset{\bar p r}{q^2 H^4 D}}{(4)} 
    =& 
    \left( 24\lambda + 4 \gamma_{H}^{(Y)}  - 3 g_1^2 - \frac{17}{3} g_2^2 \right) \coef{8}{\underset{\bar p r}{q^2 H^4 D}}{(4)}
    \nn &
    +\left[ \gamma^{(Y)}_{q} \coef{8}{\underset{}{q^2 H^4 D}}{(4)} +  \coef{8}{\underset{}{q^2 H^4 D}}{(4)} \gamma^{(Y)}_{q}  \right]_{\bar pr} 
    \nn &
    + \frac{1}{3} g_{2}^2 \left( 2\kappaPauliKinEightFour\right)\delta_{\bar pr}
    \nn &
    +\frac{1}{2} \left[   \left(- 3i\coef{8}{q^2 H^4 D}{(2)} +3i \coef{8}{q^2 H^4 D}{(3)} + 4\coef{8}{q^2 H^4 D}{(4)} \right) Y^{\dagger}_{d}Y_{d} \right.\nonumber \\ & \left. \qquad
    +  \left( 3i\coef{8}{q^2 H^4 D}{(2)} - 3i\coef{8}{q^2 H^4 D}{(3)} +4\coef{8}{q^2 H^4 D}{(4)}\right) Y^{\dagger}_{u}Y_{u}
    + \textrm{h.c.} \right]_{\bar pr}
    \,,
\end{align}
\begin{align}
    \dcoef{8}{\underset{\bar p r}{ud H^4 D}}{} =& \left(  28 \lambda  + 4\gamma_{H}^{(Y)} +\left((y_u - y_d)^2- \frac{7}{2}\right) g_1^2  - \frac{17}{6}  g_2^2  \right) \coef{8}{\underset{\bar p r}{ud H^4 D}}{}
    \nn &
    +\left[ \gamma^{(Y)}_{u} \coef{8}{\underset{}{ud H^4 D}}{(4)} +  \coef{8}{\underset{}{ud H^4 D}}{(4)} \gamma^{(Y)}_{d}  \right]_{\bar pr}
    \nn &
    +  \left[Y_{u}Y_{d}^{\dagger} \right]_{\bar pr} \left( 4\coef{8}{H^6D^2}{(1)} - 2\coef{8}{H^6D^2}{(2)} \right)  
    \nn &
    + \left[ \coef{8}{u^2 H^4 D}{} Y_{u} Y^{\dagger}_{d} - Y_{u} Y^{\dagger}_{d} \coef{8}{d^2 H^4 D}{}    \right]_{\bar pr}
    \nn &
    + \left[ \coef{8}{ud H^4 D}{} Y_{d} Y^{\dagger}_{d} + Y_{u} Y^{\dagger}_{u} \coef{8}{ud H^4 D}{}   \right]_{\bar pr}
    \,,
\end{align}
%

%%%%%%%%%%%%%%%%%%%%%%%%%%%%%%%%%%%%%%%%%%%%%%%%%%%%%%%%%%%%%%%%%%
%%%%%%%%%%%%%%%%%%%%%%%%%%%%%%%%%%%%%%%%%%%%%%%%%%%%%%%%%%%%%%%%%%

\bibliographystyle{JHEP}
\bibliography{paper-FermionRGE}

%%%%%%%%%%%%%%%%%%%%%%%%%%%%%%%%%%%%%%%%%%%%%%%%%%%%%%%%%%%%%%%%%%
\end{document}